\documentclass[preprint,eqsecnum,showpacs,aps]{revtex4}
\usepackage{graphics}
\def\beq{\begin{equation}}
\def\eeq{\end{equation}}
\def\bea{\begin{eqnarray}}
\def\eea{\end{eqnarray}}
\def\nnu{\nonumber}
\def\tst{\textstyle}
\def\muc{\multicolumn}
\def\fno#1{Fig.~\ref{#1}}
\def\cno#1{\cite{#1}}
\def\eno#1{Eq.~(\ref{#1})}
\def\Eno#1{Equation (\ref{#1})}
\def\etwo#1#2{Eqs.~(\ref{#1}) and (\ref{#2})}
\def\Etwo#1#2{Equations (\ref{#1}) and (\ref{#2})}

\def\Sno#1{Sec.~\ref{#1}}
\def\Tno#1{Table~\ref{#1}}
\def\by{\over}
\def\gtwid{\mathrel{\raise.3ex\hbox{$>$\kern-.75em\lower1ex\hbox{$\sim$}}}}
\def\ltwid{\mathrel{\raise.3ex\hbox{$<$\kern-.75em\lower1ex\hbox{$\sim$}}}}

\def\al{\alpha}

\def\gam{\gamma}
\def\dta{\delta}
\def\eps{\epsilon}
\def\tta{\theta}

\def\sig{\sigma}

\def\Gam{\Gamma}
\def\Dta{\Delta}
\def\Lam{\Lambda}

\def\Om{\Omega}
\def\Tta{\Theta}

\def\apx{\approx}
\def\ptl{\partial}

\def\hf{{1\over2}}
\def\tshf{{\tst\hf}}

\def\inmipi{\int_{-\infty}^{\infty}}
\def\inzpi{\int_0^{\infty}}
\def\grad{\nabla}

\def\part#1#2{{\ptl#1 \by \ptl#2}}

\def\lf{\left}
\def\rt{\right}

\def\ham{{\cal H}}
\def\ket#1{|#1\rangle}

\def\tran#1#2{\langle#1|#2\rangle}

\def\avg#1{\langle#1\rangle}
\def\mel#1#2#3{\langle#1|#2|#3\rangle}

\def\bb{{\bf b}}
\def\bc{{\bf c}}
\def\bh{{\bf h}}
\def\bn{{\bf n}}
\def\br{{\bf r}}

\def\bB{{\bf B}}
\def\bH{{\bf H}}

\def\bM{{\bf M}}
\def\bS{{\bf S}}

\def\xhat{{\bf{\hat x}}}
\def\yhat{{\bf{\hat y}}}
\def\zhat{{\bf{\hat z}}}
\def\nhat{{\bf{\hat n}}}

\def\itz{{\it z\ }}

\def\Fe8{Fe$_8$}
\def\Mn12{Mn$_{12}$}
\def\Mnf{Mn$_4$}
\def\up{\uparrow}
\def\dn{\downarrow}
\def\rhob{\rho_{\rm B}}
\def\hip{\ham_{i+}}
\def\him{\ham_{i-}}
\def\hp{\ham_+}
\def\hm{\ham_-}
\def\Fu{F_{\up}}
\def\Fd{F_{\dn}}
\def\Opl{\Om_+}
\def\Omi{\Om_-}
\def\Oipl{\Om_{i+}}
\def\Oimi{\Om_{i-}}
\def\epl{\eps_+}
\def\emi{\eps_-}
\def\eipl{\eps_{i+}}
\def\eimi{\eps_{i-}}

\begin{document}


\title{Incoherent Landau-Zener-St\"uckelberg Transitions in
Single-Molecule Magnets}
\author{Avinash Vijayaraghavan}
\author{Anupam Garg}
\email[e-mail address: ]{agarg@northwestern.edu}
\affiliation{Department of Physics and Astronomy, Northwestern University,
Evanston, Illinois 60208}

\date{\today}

\begin{abstract}
It is shown that in experiments on single molecule magnets (SMM's)
in which transitions between two lowest spins states are induced by
sweeping the applied magnetic field along the easy axis, the transitions
are fully incoherent. Nuclear spins and the dipolar coupling of
molecular spins are identified as the main sources of decoherence, and
the form of the decoherence is calculated. The
Landau-Zener-St\"uckelberg (LZS) process is examined in light of this
decoherence, and it is shown that the correct formula for the
spin-flip probability is better given by a more recent formula of
Kayanuma's than that of LZS. The two formulas are shown to be identical in the
limit of rapid sweeps. An approximate way of incorporating the
molecular spin dipole field into the rate equations for this process
is developed.
\end{abstract}

\pacs{75.50.Xx, 75.60.Jk, 76.20.+q, 75.45.+j}
\maketitle

\newpage
\section{Introduction}
\label{intro}

A large number of molecular solids made from organic molecules
containing magnetic ions have come to be known as single-molecule
magnets (SMM's), and their magnetization dynamics
has been studied intensively for over a decade
now~\cite{gsvbook}. The designation SMM comes about because the
intermolecular magnetic interactions are much weaker than the
intramolecular ones, yet one sees hystersis \cite{hys9597},
a phenomenon generally associated with ferromagnets in which the
spins are strongly interacting. Of special interest, and the
subject of this paper, is the study of low-temperature
quantum tunneling between the two lowest Zeeman sublevels of one
molecular spin (MS), since, then, processes such as phonon induced
excitation or relaxation do not come into play \cite{phon}, and the
dynamics is, a priori, purely quantum mechanical.

The above conclusion is strongly reinforced by experiments in which
the magnetization relaxes in the presence of a time-dependent
magnetic field which is swept through the value where the two
Zeeman levels are degenerate~\cite{ww1,ww2,ww3,wer02,edb02}.
At first sight, this constitutes a classic
Landau-Zener-St\"uckelberg (LZS) process~\cno{lzs32}, and the
data appear to confirm this idea, especially in \Fe8. The strongest
check comes from the fact that the transition probability depends
on the sweep rate over two and a half orders of magnitude in
agreement with the LZS formula \cno{ww2}. Further, the tunneling amplitude
extracted by fitting to this formula agrees with direct numerical
diagonalization of the single MS Hamiltonian. Most importantly, the
matrix element so deduced varies with a static {\it transverse\/}
magnetic field in oscillatory fashion~\cno{ww1}, as required by
the model Hamiltonian~\cno{ag93}.

It is, however, surprising that the LZS formula should be so well
obeyed, since it is derived for isolated, noninteracting spins. The
MS's in SMM's interact with many other degrees of freedom, and
anything in the environment that can distiguish between the two
tunneling states of the system will tend to suppress quantum
tunneling and act as a source of decoherence. Phonons are
an obvious such environment, but can be excluded by working at low
enough temperatures. The remaining environment is that of
the nuclear spins. These have been previously studied in connection with
magnetization tunneling in small magnetic particles ~\cno{ag9395},
and in SMM's~\cite{pro9698,sin03,sin04}.
In addition, one must also consider the other MS's. The general
picture that emerges from Refs.~\cno{pro9698,sin03,sin04} is that
nuclear spins give rise to incoherent
transitions and the other MS's spin give rise to an additional
magnetic field that must be added on to the applied field in
determining whether a given MS is at degeneracy or not.
Other authors have adopted this point of view and studied these
systems via Monte Carlo simulations~\cno{Cuc99,Fer03}.
Chapter 9 of Ref.~\cno{gsvbook} contains a good discussion of
these and related points.

Our purpose in this paper is to reexamine the decoherence from
nuclear and molecular spins, especially the latter. In \Fe8, the
dipole field from other MS's is about ten times larger than that
due to the nuclear spins, so a priori, they should be a significant
source of decoherence. It may at first sight be puzzling that the
MS's which form the ``system", can also behave as an ``environment".
The situation is analogous to how the electron-electron interaction
in metals contributes to the electrical resistivity. In a model in
which the MS's are coupled to each other, but not to any other
degrees of freedom, the {\it many-body\/} (or many-spin) wave
function of the MS's evolves coherently, yet the off-diagonal
elements of the {\it one-body\/} (one-spin) density matrix can
still decohere, i.e., decay with time. Since the magnetization is
a sum of one-spin operators, such decay is relevant to its dynamics.
Whether the model is adequate is a quantitative question depending
on whether the omitted degrees of freedom are stronger or weaker
decoherers than the ones considered. Thus, in metals at room
temperature, phonon and impurity scattering are greater contributors
to the resistivity than electron-electron scattering, and should
not be omitted in a good model. The converse is true at very low
temperatures in very pure samples (less than $\sim 1\ {\rm K}$
in potassium, for example).

The model we study is the following. Each magnetic molecule is
taken to have a total spin $S$ in its ground manifold, and to
have two easy directions, $\pm \zhat$, separated by a barrier,
$V_B$. It is assumed that other spin multiplets
can be ignored at low temperatures, so that each molecule can be
treated as a single spin of magntiude $S$. In zero external field,
an isolated MS can tunnel between the $m=\pm S$ states.
The corresponding energy splitting is denoted $\Dta$.

Next, the MS's are coupled to the nuclear spins (NS). Two broadly
different types of couplings may be distinguished.
If the magnetic ions have nuclei with nonzero magnetic moments, the
contact hyperfine interaction between an ion and its own nucleus must
be considered. The corresponding energy scale is 1--10 mK. The second
is the dipolar coupling between the MS's and other nuclear spins,
with an energy scale $E_{dn} \sim 1$~mK for close by nuclei.
(The suffixes `d' and `n' stand for `dipole' and `nuclear',
respectively.) We shall assume that $E_{dn} \gg \Dta$, as is the case
in \Fe8.

In addition, different MS's are coupled via the dipole-dipole
interaction, which is taken to have a scale $\sim E_{dm}$ for nearest
neighbours.  (The suffixes `d' and `m' stand for `dipole' and
`molecular', respectively.) There is clear separation of energy
scales: $V_B \gg E_{dm} \gg \Dta$. This is a good description of many
SMM's. In \Fe8, e.g., $V_B\sim 20$~K, $E_{dm} \sim 0.1$~K, and
$\Dta \sim 10^{-7}$--$10^{-8}$~K. Stray and dipolar magnetic fields
along $\xhat$ and $\yhat$ are unimportant since they are not large
enough to give any significant mixing of the $m = \pm S$ states with
the higher Zeeman states, and they affect $\Dta$ only weakly. Along
$\zhat$ on the other hand, such fields are very important, since they
move MS's off resonance. Under these conditions, each MS may be
replaced by a pseudospin with spin-1/2 with the $\ket{\up,\dn}$
states representing the $m=\pm S$ states of the true spin.

The plan of the paper is as follows. We calculate the decoherence from
nuclear and molecular spins in Secs.~\ref{mod1} and \ref{mod2},
respectively, pushing various details of the calculations to the
Appendices. In \Sno{comb_env}, we consider the two environments
together. In \Sno{qstat} we consider the implications of the
decoherence for the LZS process. We find that although the tunneling
is indeed incoherent, the net spin-flip probability in a single
LZS sweep is remarkably insensitive to the details of the decoherence
mechanism. In a simple model where the dipole field due to the
other MS's is omitted, the probability turns out to be given exactly
by Kayanuma's formula for a spin coupled to an oscillator bath in
the strong damping limit~\cno{Kaya}. In the limit of high field sweep
rate this formula agrees precisely with the LZS formula. This explains
why the experiments appear to be in accord with the LZS scenario.
We also consider a better approximation where the dipole field is
included in a macroscopically averaged way. This approximation improves
the agreement with the experiments by Wernsdorfer et al.~\cno{ww2,ww3}.

\section{Model for nuclear spin environment}
\label{mod1}

As our first model, we consider a single molecular spin interacting
with the nuclear spins via the dipolar coupling. Hyperfine and transferred
hyperfine interactions are not explicitly included, although in the
end they are unlikely to have qualitatively different effects, and
only to lead to a modification of the parameter $W$ introduced below.
We assume that all nuclear spins have spin 1/2, and neglect the local
magnetic field $H_{\rm loc}$ at the nuclear site. This is a good
assumption if $H_{\rm loc} \ll k_B T/\mu_n$, where $\mu_n$ is the
nuclear magnetic moment. This is indeed so since $k_B T/\mu_n \sim 10$~T
at 10 mK. The dipolar coupling between nuclear spins can be neglected
for the same reason. With these assumptions, our Hamiltonian is
\beq
\ham_{mn}
  = \hf(\Dta\sig_{0x} + \eps \sig_{0z})
    +\sum_i {E_{dn} a^3 \by r^3_{0i}}
           [ \sig_{0z}\sig_{iz}
              - 3\sig_{0z} \cos \tta_{0i} \vec\sig_i\cdot{\hat\br}_{0i}].
           \label{hmn}
\eeq
Here, $i$ labels the different nuclear spins, $\vec\sig_0$ and
$\vec\sig_i$ denote the Pauli spin matrices for the MS and the $i$th
nuclear spin, $\br_{0i}$ is the position of the $i$th NS relative to
the MS, $r_{0i} = |\br_{0i}|$, and
$\cos\tta_{0i} = \zhat\cdot\br_{0i}/r_{0i}$. Further, $a$ is the
characteristic distance from the MS to the nearest NS. We expect
$a \sim$ 1-2 \AA\  for any SMM. Finally, we have included an energy bias
$\eps$ between the $\ket\up$ and $\ket\dn$ states of the MS, which
could arise from an external magnetic field. The suffixes in
$\ham_{mn}$ stand for `molecular' and `nuclear'.

We now suppose that at time $t=0$ the MS is in the state $\ket{\!\up}$,
and that every NS is in a completely disordered state described by the
density matrix $1/2$. Again, this assumption is well justified at
the temperatures at which experiments have been carried out so far. The
quantity of interest is the probability, $P(t)$, that the MS will
be in the state $\ket{\dn}$ irrespective of the NS state.

Even for this simple model, an exact calculation of $P(t)$ is not
possible (but see below).
We therefore turn to the approximate methods described
in Sec. III A--D of Ref.~\cno{sixman}. We cannot assume that the
damping is weak, or that the NS's are fast compared to the MS's.
A ``golden rule" approach is still fruitful, however, as $\Dta$ is the
smallest energy scale in the problem. Moreover, the validity of this
approach can be self-consistently checked. Second-order perturbation
theory yields
\beq
P(t) = {\Dta^2 \by 4} \int_0^t dt_1 \int_0^t dt_2\,
                  e^{i\eps(t_1 - t_2)} \prod_i F_i(t_1,t_2),
     \label{gold1}
\eeq
where
\beq
F_i(t_1,t_2)
  = \hf {\rm Tr}_i
       \left[e^{i\hip t_1} e^{i\him(t_1 - t_2)}
                                    e^{-i\hip t_2}\right],
    \label{Fni}
\eeq
with
\beq
\ham_{i\pm}
   = \pm {E_{dn} a^3 \by r^3_{0i}}
           [\sig_{iz}
              - 3 \cos\tta_{0i} \vec\sig_i\cdot{\hat\br}_{0i}].
        \label{hni}
\eeq
The quantity $F_i$ is the contribution of the $i$th environmental
spin to Feynman's influence functional evaluated for
a particular pair of forward and backward paths of the ``system"
spin, namely, that in which this spin flips from up to down at
time $t_1$ on the forward path, and time $t_2$ on the backward
path. We therefore refer to $F_i$ as the (environmental) influence
factor or function.

The trace in \eno{Fni} is easy to evaluate. Defining
\beq
t_{12} = t_1 - t_2.
\eeq
$x_{0i} = \br_{0i}\cdot\xhat$, etc., and the vector
\beq
\bh_i = {E_{dn}a^3 \by r_{0i}^5}
          (-3z_{0i} x_{0i}, -3z_{0i} y_{0i}, r_{0i}^2 - 3z^2_{0i}),
\eeq
we have
\beq
F_i(t_1, t_2) = \cos 2 h_i t_{12}.
\eeq
Now,
\beq
h_i = {E_{dn} a^3 \by r_{0i}^3} (1 + 3\cos^2\tta_{0i})^{1/2},
\eeq
so $h_i \sim E_{dn}$ for the nearest NS, and drops as $1/r^3$ for
more distant ones. Thus, for
$t_{12} \gtwid E_{dn}^{-1}$ the different $F_i$'s have
random signs, and since they can not exceed 1 {in} magnitude, they
essentially multiply out to zero. We conclude that phase coherence
is lost on the time scale $t_c \sim E_{dn}^{-1}$, and for
$t \gg t_c$, we get incoherent tunneling. For such times, we can
approximate
\beq
\prod_i F_i(t_1,t_2)
   \simeq \exp \bigl(- 2\sum_i h_i^2 t_{12}^2 \bigr).
\eeq
Further, in the double integral in \eno{gold1}, we may introduce
sum and difference variables $\bar t = (t_1 + t_2)/2$ and
$\tau = t_{12}$. The integral over $\tau$ is essentially independent
of $\bar t$, and its limits may be extended to $\pm\infty$. The
$\bar t$ integral then gives an overall factor of $t$, yielding
\beq
P(t) \simeq \Gam_n t,   \label{fgr}
\eeq
where, with,
\beq
W^2 = 4\sum_i h_i^2, \label{defW}
\eeq
\beq
\Gam_n = {1\by 4} \Dta^2
          \inmipi d\tau e^{i\eps\tau} e^{-\hf W^2 \tau^2}
       = {\sqrt{2\pi} \by 4} {\Dta^2 \by W} e^{-\eps^2/2 W^2}.
\eeq

We may estimate $W$ by replacing the sum in \eno{defW} by an
integral, taking a uniform density of nuclear spins equal to
$1/a^3$ outside a sphere of radius $a$. Since
\beq
h_i^2 = {E_{dn}^2 a^6 \by r_{0i}^6}
            (1 + 3\cos^2\tta_{0i}),
\eeq
\bea
W^2 &\simeq& 4E_{dn}^2 a^6
       \int_{r > a} {d^3 r \by a^3} (1 + 3\cos^2\tta) {1\by r^6} \\
    &=& {32\pi \by 3} E_{dn}^2.
\eea
In fact, the integral estimates the contribution of the nearest
neighbors rather poorly, and for the simple, body-centered, and
face-centered cubic lattices, the number multiplying $E_{dn}^2$
is 67.2, 98.0, and 116, respectively~\cite{am}. Thus, in order
of magnitude, we may take $W \simeq 10 E_{dn}$ for any magnetic
molecular solid. It should be noted that for a fixed bias $\eps$,
the rate $\Gam_n$ goes up with increasing $E_{dn}$, as long as
$\eps > W$. The converse is true for the very small number of
MS's on which the bias is small, $\eps < W$.

The result (\ref{fgr}) is essentially a Fermi golden rule rate,
and is limited to $t \ll \Gam_n^{-1}$. For longer times, a formal
answer can be obtained as follows \cno{dob}. We can write
\beq
\ham_{mn} = \hf {\vec\Lam}\cdot{\vec\sig}_0,
\eeq
where
\beq
{\vec\Lam} = \Dta\xhat
               + (\eps + 2\sum_i \bh_i \cdot {\vec\sig}_i) \zhat.
\eeq
Thus, ${\vec\Lam}$ is an operator with respect to the bath spins.
With the understanding that these must be traced over, we get
\beq
\mel{\dn}{e^{-i\ham_{mn}t}}{\up}
       = -{i\Dta\by \Lam} \sin\tshf \Lam t.
\eeq
Thus,
\beq
P(t) = \Dta^2 \prod_i \hf {\rm tr}_i
             \lf({1\by \Lam^2} \sin^2 {\Lam t \by 2} \rt),
\eeq
where ${\rm tr}_i$ indicates a trace over the $i$th NS. To 
perform this trace we take the quantization axis for it to be
parallel to $\bh_i$. This means that the variable
\beq
B_n = 2\sum_i h_i s_i
\eeq
takes on all possible values obtained by letting each $s_i$ be $+1$
or $-1$ independently \cno{B_clar}. That is to say, $B_n$ is a stochastic
variable with some probability distribution, $P(B_n)$, and the spin
flip probability is obtained by averaging over this distribution:
\beq
P(t) = \inmipi {\Dta^2 \by \Lam^2} \sin^2(\tshf\Lam t) P(B_n) \, dB_n,
\eeq
with
\beq
\Lam = \lf(\Dta^2 + (\eps + B_n)^2 \rt)^{1/2}.
\eeq

To proceed further, we need the form of $P(B_n)$. We find this
approximately by arguing that because of the law of large numbers
$B_n$ is a Gaussian with a variance $W^2$, i.e.,
\beq
P(B_n) = \lf( {1\by 2\pi W^2} \rt)^{1/2} e^{-B_n^2/2W^2}.
              \label{B_Gaus}
\eeq
We can do somewhat better by looking at the moments of $B_n$. We
clearly have $\avg{B_n^2} = W^2$, but
\beq
\avg{B_n^4} = 3{\avg{B_n^2}}^2 - 32 \sum_i h_i^4.
\eeq
Thus the fourth moment is less than what it is for a Gaussian
(negative kurtosis),
and the distribution has less weight in the wings than a Gaussian.
We shall see that that the detailed form of $P(B_n)$ is not too
important, and for our purposes, \eno{B_Gaus} is good enough.

For $W^{-1} \ll t \ll \Dta^{-1}$, we may evaluate $P(t)$ by replacing
$\Lam$ by $(B_n+\eps)$. (This replacement is no longer valid when
$\Dta t \gtwid 1$, for then the phase of $\sin^2(\Lam t/2)$ is
significantly altered by throwing away $\Dta$.) Then by the
usual textbook argument for Fermi's golden rule,
\beq
{\sin^2 \bigl((B_n+\eps)t/2 \bigr) \by (B_n+\eps)^2}
      = {2\pi t\by 4} \dta(B_n+\eps). 
\eeq
The integral for $P(t)$ is then trivial, and yields
\beq
P(t) = {\sqrt{2\pi} \by 4} {\Dta^2 \by W} e^{-\eps^2/2 W^2}t,
\eeq
which is the same as before.

For $\Dta t \gtwid 1$, the integral is dominated by
$B_n \apx -\eps$, and we may put $B_n = -\eps$ in the Gaussian factor,
yielding
\bea
P(t) &=& {\Dta^2 \by \sqrt{8\pi} W}
            e^{-\eps^2/2W^2}
               \inmipi {1 - \cos(\sqrt{\Dta^2 + b^2} t)
                   \by \Dta^2 + b^2} db \nnu \\
     &=& \sqrt{\pi \by 8} {\Dta \by W}
            e^{-\eps^2/2W^2}
              \lf( 1 - \int_{\Dta t}^{\infty} J_0(z) dz \rt),
\eea
where $b = B_n+\eps$, and we used Ref.~\cno{GR} in the last step.
Using the asymptotic behaviour of the Bessel function, we find that
for $\Dta t \gg 1$,
\beq
P(t) \apx \sqrt{\pi \by 8}
          {\Dta \by W}
            e^{-\eps^2/2W^2}
             \lf[1 - \sqrt{2\by \pi\Dta t}
                        \sin\Bigl(\Dta t - {\pi\by 4}\Bigr) \rt].
\eeq
The important point is that even for $\eps = 0$, the nuclear spin
environment impedes the spin from flipping appreciably, and the
net flip probability is only of order $\Dta/W$.

\section{Model for molecular spin environment}
\label{mod2}
For our second model, we consider only the dipolar coupling between
MS's, and ignore the nuclear spins altogether. Let us denote the energy
scale of the mutual dipole-dipole interaction between MS's by
$E_{dm}$ [see \eno{def_Ki2} below for the exact definition].
Since $E_{dm} \gg E_{dn}$, we may a priori expect
decoherence by the mutual interaction to be much greater than that
by the interaction with NS's. This model is studied in an attempt to
investigate this point.

In terms of the Pauli matrices, the Hamiltonian for interacting MS's
can be written as
\beq
\ham_C = \hf\sum_i (\Dta\sig_{ix} + \eps_i \sig_{iz})
           + \hf \sum_{i < j} K_{ij} \sig_{iz}\sig_{jz}.
      \label{ham_coup}
\eeq
Here $i$ and $j$ label the different spins, $x$ and $z$ denote
the axes, $\eps_i$ is the bias field on spin $i$ that moves it
off-resonance, and $K_{ij}$ is the dipolar coupling.

Let us now focus on one MS, which we shall
call the system, and label it with a suffix $0$. This is prepared in
the $\ket\up$ state at time $t=0$, and the other spins, which we call
the bath, are prepared in a density matrix $\rhob$.  Let $P(t)$
denote the probability that the system spin is in the state $\ket{\dn}$
at a later time $t$ irrespective of the state of the bath. For
an isolated spin, $P(t) = \sin^2(\Dta t/2)$.
If decoherence is weak, we expect the oscillations to be weakly
damped, and if it is strong, we expect a decay without any
oscillation. Indeed, these qualitative behaviours define what we
mean by weak and strong decoherence. Since the dipole interaction
is long-ranged, we anticipate that the decoherence might depend
on the spatial position of spin {0} in the sample, especially
if $\rhob$ corresponds to a fully or nearly fully polarized bath,
but otherwise there is nothing special about its choice.

The calculation of $P(t)$ for the model (\ref{ham_coup}) appears
daunting because of the couplings between the bath spins. We therefore
consider a simpler model
\beq
\ham_{mm} = \hf (\Dta\sig_{0x} + \eps \sig_{0z})
           + \hf\sum_{i\ne 0} (\Dta\sig_{ix} + \eps_i \sig_{iz})
           + \hf \sum_{i \ne 0} K_i \sig_{0z}\sig_{iz}.
      \label{ham_dec}
\eeq
(Both suffixes in $\ham_{mm}$ stand for `molecular'.)
The dipolar couplings between the bath spins are now replaced
by a distribution of dipole fields by treating the bias energies
$\eps_i$ as independent random
variables, distributed on the scale $E_{dm}$. The calculation of
$P(t)$ should include an ensemble average over this distribution.
The coupling $K_i$ between spin {0} and spin $i$ of the bath
is, however, retained as is, and is, explicitly,
\beq
K_i = {2 E_{dm} a^3 \by r_{0i}^3} (1-3\cos\tta_{0i}^2).
    \label{def_Ki2}
\eeq
Here, $a$ is the nearest neighbour distance, $r_{0i}$ is the
distance from spin $0$ to spin $i$, and $\tta_{0i}$ is the angle
the line joining them makes with the \itz axis. Finally, $\eps$
is an additional bias on spin {0}, due to an external field,
for example.

For purposes of explicit calculation, we shall take the probability
density of the biases $\eps_i$ to be Gaussian,
\beq
f(\eps) = {1\by\sqrt{2\pi E_b^2}} e^{-\eps^2/2E_b^2},
    \label{dist_eps}
\eeq
where $E_b \sim E_{dm}$. Dipolar field distributions in \Fe8 have
been measured by Ohm, Sangregorio, and Paulsen \cite{ohm}, and by
Wernsdorfer et al.~\cite{ww4}. They have also been inferred from
linewidth measurements in optical spectroscopy by Mukhin et
al.~\cite{Mukhin01}. The assumption of a Gaussian form is
consistent with these measurements. Berkov has given theoretical and
Monte Carlo arguments for a Gaussian distribution in a system of
dense interacting dipoles~\cno{Ber96}. We shall see, nevertheless,
that the detailed form of this distribution is not physically
important for us.

Even the model (\ref{ham_dec}) cannot be treated exactly. It
is again seen that the weak coupling approximation is totally
invalid, and adiabatic renormalization is inapplicable since the
bath and system spins move on the same time scale. The
golden rule is still good, however. Second-order
perturbation theory in $\Dta$ yields
\beq
P(t) = {\Dta^2 \by 4} \int_0^t dt_1 \int_0^t dt_2\,
                  e^{i\eps(t_1 - t_2)} F,
     \label{gold}
\eeq
where
\beq
F = {\rm Tr}_{\rm B} \left[\rhob
         \prod_i e^{i\hip t_1}
                    e^{-i\him(t_1 - t_2)}
                      e^{-i\hip t_2}\right],
\eeq
with
\beq
\ham_{i\pm} = \hf \Big( \Dta\sig_{ix} + (\eps_i \pm K_i)
                                 \sig_{iz}
                                      \Big).
\eeq

The choice of $\rhob$ demands some care. It would now be incorrect to
take $\rhob = 2^{-N_m}$, where $N_m$ is the number of MS's since these
spins do not equilibrate between the $\ket{\up}$ and $\ket{\dn}$
states on a time scale short compared to $\Dta^{-1}$. Instead we choose
each spin to be in a definite state, either $\ket\up$ or $\ket\dn$.
(In the language of statistical mechanics, the bath is in a state of
quenched disorder.)
This then means that in principle we have to calculate the second order
influence function for every configuration of MS's separately. In
practice, this is not so, and we shall see that the functions for
up and down spins differ only by phases. When these phases are added
together for all the MS's in the bath, they will reproduce exactly the
effect of the local dipole field at spin 0. This field is dependent
on the MS configuration, but except for special configurations such as
all or nearly all MS's polarized in the same direction, we can treat it
statistically as a field with an rms value of order $E_{dm}$.

In equations, the above means that if we specify the spin configuration
by giving $s_i = \avg{\sig_{iz}} = \pm 1$, then
\beq
\rhob = \prod_i \rho_i\ ; \quad
      \rho_i = \hf(1 + s_i \sig_{iz}).
\eeq
Accordingly, $F$ factorizes into a product of factors, one for each
bath MS. If the $i$th spin is ``up", this factor is
\beq
F_i = \mel{\up}
                     {e^{i\hip t_1} e^{-i\him(t_1 - t_2)}
                             e^{-i\hip t_2}}{\up}.
   \label{infn_up}
\eeq
If the spin is ``down", $F_i$ is given by the expectation value of the
same operator in the $\ket\dn$ state. The calculation of these
influence factors is lengthy, and is presented in Appendix
\ref{F_calc}. We find that
\beq
F_i \simeq e^{is_i K_i t_{12}} (1 - \eta_i),
\eeq
with $\eta_i$ given by \eno{mis} with the addition of a suffix $i$
to $K$ and $\Om_{\pm}$, $\bar t = (t_1 + t_2)/2$, and
$t_{12} = t_1 - t_2$.

We call the quantity $\eta_i$ the {\it mismatch\/}, since it arises
from a difference in the time evolution of the $i$th environmental
spin in response to different paths taken by the system spin. The
derivation in Appendix \ref{F_calc} shows that $0 \le \eta_i \ll 1$,
vanishing only when $t_{12} = 0$ \cite{somz}.
Hence we may put $1-\eta_i \apx e^{-\eta_i}$, leading to
\beq
P(t) = {\Dta^2 \by 4} \int_0^t dt_1 \int_0^t dt_2\,
                  e^{i\eps_T t_{12}} e^{-\sum_i \eta_i},
     \label{gold2}
\eeq
where
\beq
\eps_T = \eps + \sum_i K_i s_i.
\eeq
This is the total bias that the spin at 0 sees including the dipole
field of the other MS's. Its value is of order $E_{dm}$ except for
special spin configurations.

We show in Appendix \ref{eta_estimate} that for
$|t_{12}| \gg E_{dm}^{-1}$, and
$\Dta^{-1} \ll \bar t \ll E_{dn}^2/\Dta$,
\beq
\sum_i \eta_i \simeq \gam_m \Dta |t_{12}|,
    \label{sum_eta}
\eeq
where $\gam_m$ is a constant of order unity. We have also evaluated
this sum numerically, as described in Appendix \ref{eta_num}. This
work shows that the form (\ref{sum_eta}) is good even for
$\Dta |t_{12}| \sim 1$. Employing it in \eno{gold2}, we get
\beq
P(t) = {\Dta^2 \by 2}
        {\rm Re}\lf[ {t \by (\gam_m \Dta - i\eps_T)}
                       - {1 - e^{-(\gam_m \Dta - i\eps_T) t}
                            \by
                              (\gam_m \Dta - i\eps_T)^2} \rt].
\eeq
Thus, $P(t)$ displays damped oscillations about a slowly rising mean.
The time scale of the decoherence is $\Dta^{-1}$, which is comparable
to the time scale of the oscillations when the total bias, $\eps_T$, is
zero. The amplitude of the oscillations is $\sim\Dta^2/\eps_T^2$ if
the bias is large. For $t \gg \Dta^{-1}$, we obtain
\beq
P(t) \simeq \Gam_m t,
\eeq
with
\beq
\Gam_m = \hf {\gam_m \Dta^3 \by \gam_m^2 \Dta^2 + \eps_T^2}.
\eeq
%
This quantity may be interpreted as an average {\it rate\/} at which
the spin flips.
If the net bias is large ($\gg \Dta$), this rate is
$\gam_m \Dta^3/2 \eps_T^2$, while if the bias is zero, it is
much larger, $\Dta/2\gam_m$. (The amplitude of the oscillations is
also very small when the bias is large.)
It is interesting that the
zero-bias rate is proportional to $\Dta$ and not to $\Dta^2$ as might
be expected from a naive application of the golden rule; this is
because the decoherence time scale is also set by $\Dta$.

\section{Combined nuclear and molecular spin environments}
\label{comb_env}

Let us now consider both environments together. The combined influence
factor is the product of the influence factors for each separate
environment, leading to
\beq
P(t) = {\Dta^2 \by 4} \int_0^t dt_1 \int_0^t dt_2\,
                  e^{i\eps_T t_{12}} e^{-\gam_m \Dta |t_{12}|}
                     e^{-W^2 t_{12}^2 /2}.
     \label{gold3}
\eeq
If, as is generally the case, $W \sim E_{dn} \gg \Dta$, the integrals
may be evaluated as in \Sno{mod1}. We once again get
$P(t) \simeq \Gam t$, with
\beq
\Gam = {\Dta^2 \by 4} \inmipi dt\, e^{i\eps_T t}
                            e^{-\gam \Dta |t|} e^{-W^2 t^2/2}.
    \label{Gam_comb}
\eeq
In general this integral leads to an error function, but if
$E_{dn} \gg \Dta$, it simplifies, and we get
\beq
\Gam = {\sqrt{2\pi} \by 4} {\Dta^2 \by W} e^{-\eps_T^2/2W^2}.
\eeq
This is of the same form as $\Gam_n$, and the main effect of the
molecular spins is to change the bias field.
\section{Quasistatic model of field sweeps}
\label{qstat}

We have seen in the previous sections that the molecular
spin relaxes incoherently from $\ket\up$ to $\ket\dn$. There may
in addition be some vestige of the coherent oscillations, but these
decay because of the coupling to nuclear spins and to other
molecular spins. The decay time scales due to these two couplings
are $E_{dn}^{-1}$ and $\Dta^{-1}$ respectively, and the former is
the relevant one since it is so much shorter. If the externally
applied  field is swept slowly enough that the bias on any one spin
changes by much less than $E_{dn}$ in a time $E_{dn}^{-1}$, that is,
if $\dot\eps_T \ll E_{dn}^2$, then it is a good
approximation to neglect the off-diagonal elements of the density
matrix, and to write simple rate equations for the diagonal elements.
If we denote the probability for a particular molecular spin to be
in the $\ket\up$ or $\ket\dn$ states by $p_{\up}$ and $p_{\dn}$,
we have,
\beq
{d p_{\up} \by dt}
   = \Gam (\eps_T(t)) (p_{\dn} - p_{\up})
   = \Gam (\eps_T(t)) (1 - 2 p_{\up}).
    \label{rate_eqn}
\eeq
where the rate $\Gam$ has been allowed to vary with time through
its dependence on the bias. Let the spin state be $\ket\dn$ at
$t = -\infty$. Then, \eno{rate_eqn} is easily integrated to yield
\beq
p_{\up}(t)
 = \hf \lf[ 1 - \exp\lf(-2\int_{-\infty}^t \Gam(\eps_T(t')) dt' \rt)\rt].
     \label{rate_soln}
\eeq
In particular, the probability for the spin to flip is given by
\beq
p_f \equiv p_{\up}(\infty)
 = \hf \lf[ 1 - \exp\lf(-2\inmipi \Gam(\eps_T(t')) dt' \rt)\rt].
   \label{p_flip}
\eeq

It is interesting to analyze the spin flip probabilty neglecting the
contribution of the other molecular spins to the bias. That is, we
take $\eps_T(t)$ to be $\eps_a(t)$, the applied bias field. Further, as
in the standard LZS protocol, we take $\dot\eps_a$ to be a constant.
Such an analysis would be directly applicable to a situation in which
the molecular spins were very dilute and $E_{dm}$ was smaller than
$E_{dn}$. Since we chose $p_{\up}(-\infty) = 0$, we must take the
bias field to be swept from large positive to large negative values
and the integral in \eno{p_flip} becomes
\beq
\inmipi \Gam(\eps_a(t)) \,dt
    = {1\by |\dot\eps_a|} \inmipi \Gam(\eps_a) \,d\eps_a.
\eeq
\Eno{Gam_comb} now yields (writing $\eps_a$ for $\eps_T$)
\bea
\inmipi\Gam(\eps_a) \,d\eps_a
  &=& {\Dta^2 \by 4} \inmipi d\eps_a \inmipi dt\, e^{i\eps_a t}
                            e^{-\gam \Dta |t|} e^{-W^2 t^2/2} \nnu\\
  &=& {\Dta^2 \by 4} \inmipi dt\, 
             e^{-\gam \Dta |t|} e^{-W^2 t^2/2}
              2\pi \dta(t) \nnu\\
  &=& {\pi \Dta^2 \by 2}.
\eea
Hence,
\beq
p_f = \hf (1 - e^{- \pi \Dta^2/|\dot\eps_a|}).
    \label{pf_qs}
\eeq
This is the same as Kayanuma's result \cno{Kaya} for a spin
coupled to an oscillator bath in the strong damping limit. Our
derivation shows that this result is valid more generally whenever
the transitions are so incoherent as to allow for rate equations.
The striking fact that the details of the decoherence mechanism
drop out of the final result can also be understood. If the
decoherence is large, the rate $\Gam$ is small, but the spin
can flip over a larger energy interval around the crossing, i.e.,
over a larger range of bias energy. For pure nuclear spin
decoherence, $\Gam \sim \Dta^2/W$, but the crossing region is broadened
to a width $\sim W$. For pure molecular spin decoherence,
$\Gam \sim \Dta$, and the crossing region is also of width $\sim\Dta$.

The quasistatic result (\ref{pf_qs}) should be compared with the
LZS spin-flip probability,
\beq
p_{f,{\rm LZS}} = (1 - e^{-\pi \Dta^2/2 |\dot\eps_a|}).
 \label{p_LZS}
\eeq
In the fast-sweep limit, i.e., with $|\dot\eps_a| \gg \Dta^2$,
$p_f \ll 1$, and the two results are identical,
\beq
p_f = p_{f,{\rm LZS}} \simeq {\pi \Dta^2 \by 2 |\dot\eps_a|}.
      \label{Kaya_LZS}
\eeq
This remarkable result has very interesting implications for the
experiments by Wernsdorfer and colleagues \cno{ww1,ww2,ww3}. It
has always been a surprise that the data in these experiments agree
with the LZS formula, even in the fast sweep limit. After all, the LZS
formula is derived for a single noninteracting spin, and the spins in
\Fe8 are not noninteracting, and are subject to strong and
rapidly fluctuating fields from the NS's and possibly the MS's.
Indeed, it is the systematics of the agreement with the LZS formula
that has been used to argue that one can extract the underlying
tunneling matrix element from the incoherent relaxation of the net
magnetization in a swept external field. \Eno{Kaya_LZS} provides an
explanation of this fact. It also means, in a stroke of
luck, that the analysis of Ref.~\cno{kg07} continues to be valid.

In the slow-sweep limit, on the other hand, $p_{f,{\rm LZS}} \simeq 1$,
while $p_f \simeq 1/2$. This means that if we continue to infer a
tunneling matrix element, $\Dta_{\rm inf}$, by fitting the flip
probability to an LZS form, we have
\beq
\Dta^2_{\rm inf} (\dot\eps)
   = -{2 \dot\eps \by \pi}
           \ln \lf( {1 + e^{-\pi \Dta^2/{|\dot\eps|}}} \by 2 \rt).
\eeq
We plot this in \fno{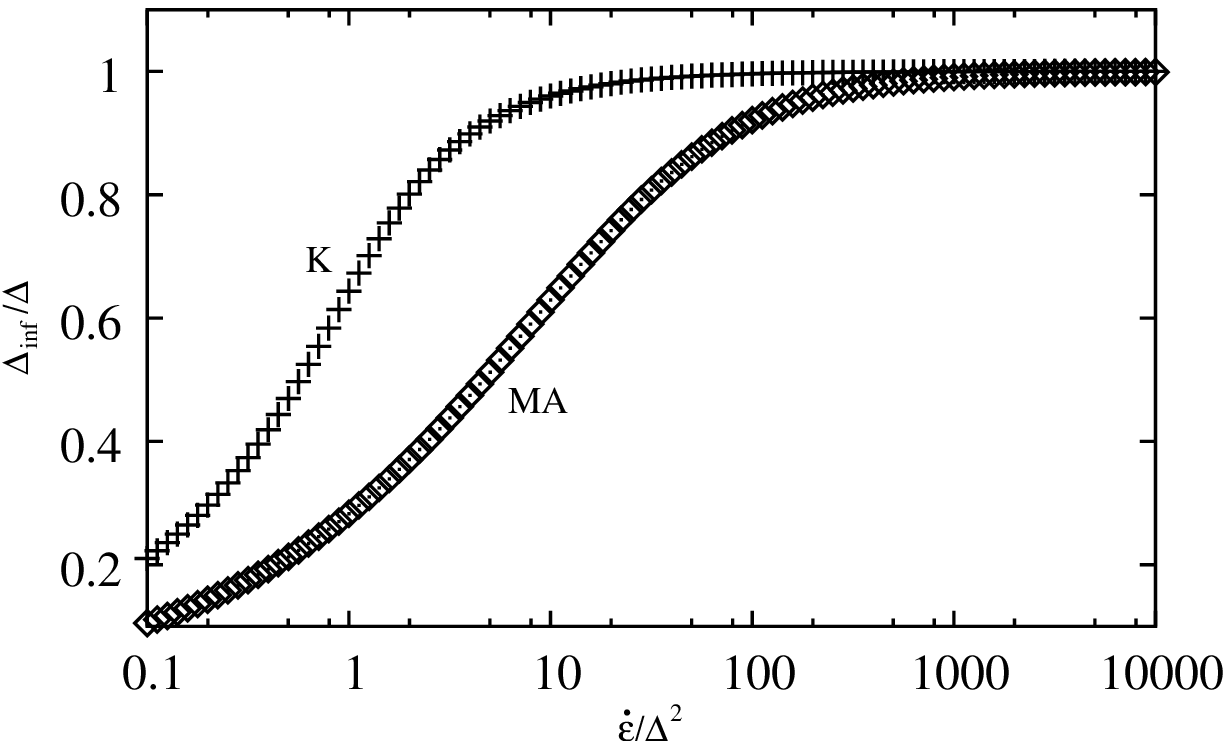}, which should be compared with
Fig. 7 of Ref.~\cno{ww2}. Although our plot is qualitatively similar,
it does not agree in detail. In particular, the experimentally
inferred splitting drops more rapidly with $\dot\eps$ once
$\dot\eps \ltwid 50 \Dta^2$ than our model shows. Neverthless, the
general trend indicates that we have captured some of the essential
physics. On the other hand, this simple formula does not contain the
experimentally seen dependence of the splitting as inferred from the
fast-sweep data on the nuclear spin coupling \cno{ww3}.

To prevent misunderstanding, we note here that we have described
the sweep as fast or slow depending on the ratio
$|\dot\eps_a|/\Dta^2$. However, because $\Dta\ll E_{dn}$, even if
$|\dot\eps_a| \gg \Dta^2$, it is possible to satisfy
$|\dot\eps_a| \ll E_{dn}^2$, the condition for the quasistatic
treatment to apply.

Let us now ask how to include the effect of the mutual dipole field
as the external magnetic field is swept. The picture that emerges
is that each MS flips at a rate that depends on the bias field seen
by it. If at time $t$, the MS configuration is set of Ising spin
variables $\{s_i\}$, and the net bias on the $i$th spin is
$\eps_{iT}$, then at a short time $\Dta t$ later,
\beq
s_i \to -s_i {\rm \ with\ probability\ } \Gam(\eps_{iT}(t)) \Dta t.
\eeq
The spins and the dipole fields then become a complicated coupled
stochastic process.
As noted in Ref.~\cno{gsvbook}, this is a Glauber process with
the difference that the flipping rate depends on the long ranged
dipole field. Monte Carlo studies of such processes have been
performed by Cuccoli et al.~\cno{Cuc99}, and by Fernandez and
Alonso~\cno{Fer03}. It would be interesting to conduct similar
studies in a swept field with the rates found by us.

Here, we consider a simpler way to incorporate the dipole field
in the rate equation (\eno{rate_eqn}) in an average way that
ignores its site to site variation, through
the macroscopic demagnetization field. To forestall confusion,
it pays to recall the distinction between $\bB$, $\bH$, and the
contribution of the demagnetization field to the latter. We work
in the Gaussian system of units. Let $\bH_a$ be the applied
magnetic field, i.e., the field that a solenoid wound around the
sample would produce if the sample were not there. Let $\bM$
be the magnetization, i.e., the magnetic dipole moment density,
and let $\bH_{\rm demag}$ be the demagnetization field, i.e., the
field produced by a volume charge density $\grad\cdot\bM$
and a surface charge density $\bM\cdot\nhat$, where $\nhat$ is the
outward normal at the surface of the sample. The field to which
an individual MS responds is the induction
\beq
\bB = \bH_a + 4\pi\bM + \bH_{\rm demag},
\eeq
through a term in the Hamiltonian,
\beq
\ham_{\rm bias} = -g\mu_B \bS^{\rm op}\cdot\bB.
\eeq
Here, $\bS^{\rm op}$ is the operator for the total spin of the
molecule in question. Since, as we have argued, the MS's behave
as essentially classical variables with only a {\it z\/} component,
the total bias on this MS is
\beq
\eps_T = - 2g\mu_B S B_z.
\eeq
Henceforth we will write just $H_a$ and $M$ for $H_{a,z}$ and $M_z$.
For simplicity we will ignore the spatial inhomogeneity of $\bM$ and
$\bH_{\rm demag}$, as well as the tensorial character of their
proportionality, and write
\beq
4\pi\bM + \bH_{\rm demag} = \al \bM.
\eeq
The constant $\al$ is shape dependent: it would be $8\pi/3$ for a
perfectly uniformly magnetized sphere, $4\pi$ for a thin long rod
parallel to $\bH_a$, and $0$ for a thin flat disc normal to
$\bH_a$.

With the above definitions, the bias is given by
\beq
\eps_T = -2 g \mu_B S
           \bigl( H_a - \al n g\mu_B S (1-2p_{\up})\bigr),
\eeq
where $n$ is the number density of MS's. Hence,
$n (g\mu_B S)^2 \sim E_{dm}$. Let us again take
$p_{\up}(-\infty) = 0$ and $H_a < 0$ at $t \to -\infty$, so
$\dot\eps_a < 0$. Adjusting the zero of time, and absorbing another
constant of order unity in $\al$, we get
\beq
\eps_T(t) = \dot\eps_a t - 4\al E_{dm} p_{\up}(t),
\eeq
and feed this into the rate equation (\ref{rate_eqn}). The resulting
differential equation for $p_{\up}$ is
\beq
{dp_{\up} \by du}
  = {\sqrt{2\pi} \by 4}{\Dta^2 \by |\dot\eps_a|} (1 - 2p_{\up})
      \exp\lf[ -\hf \lf( u + {4\al E_{dm}\by W} p_{\up} \rt)^2 \rt],
 \label{pf_MA}
\eeq
where $u = |\dot\eps_a| t/W$. This equation can also be formally
solved as before, by treating $\Gam(\eps_T(t))$ as a known function
of time.  The solution is then again given by \eno{rate_soln}, but
since $\eps_T(t)$ depends on $p_{\up}(t)$, it is now in the form of
an integral equation. We have found it simpler to integrate the
differential equation numerically for different values of $\dot\eps$.
The results are shown in \fno{Dta_inf.eps}. The qualitative agreement
with Ref.~\cno{ww2} is improved, although we cannot make a direct
comparison because of uncertainty in the ratio $\al E_{dm}/W$.

Finally, let us return to the point that for slow sweeps,
$p_{\up}(\infty) \simeq 1$ for coherent LZS sweeps, while
$p_{\up}(\infty) \simeq \hf$ for incoherent sweeps, both from
Kayanuma's formula (\ref{pf_qs}) or the formula (\ref{pf_MA}) which
includes MS dipolar fields~\cno{merci}. This means that starting from a sample
with a saturated magnetization $-M_0$, we are arguing that the final
magnetization will be $0$ and not $M_0$ if the field is swept slowly.
And indeed, studies of the \Mnf~\cno{werns05} and \Mn12 wheel
SMM's~\cno{werns08} show just such behaviour. In Ref.~\cno{werns05},
it is found that the final magnetization is zero for slow sweeps (see
Fig. 8a, 8b, and 8c there), and is fully reversed only for ultra slow
sweeps (Fig. 8d and 8h). For sweep rates in between, and for inverse
LZS sweeps (Fig. 8c,8f, and 8g), the final magnetization is not zero,
but is not completely reversed either. (See also Fig. 1a of
Ref.~\cno{werns08}. This supports the conclusion
that the transitions are incoherent. It also means that to fully
explain the ultra-slow sweep and the inverse LZS sweep data, one must
have a mechanism for the spin to relax from the higher energy state to
the lower energy one even (but not vice versa) when the bias is much
larger than $E_{dn}$. One possibility is to have a second order
Fermi golden rule process in which (assuming the spin is 10),
the spin tunnels from the $m=-10$ to an $m=9$ or $m=8$ virtual state
followed by a transition to the $m=10$ state with the emission of a
phonon. We shall address this issue further in a separate publication.

\section{Discussion}
\label{Disc}

We have considered the transitions in a swept field in the presence
of nuclear and molecular spin decoherence. Our qualitative conclusions
regarding the former are in accord with those of
Refs.~\cno{pro9698,sin03,sin04},
but the quantitative form of the decoherence is
different. Similarly, with regard to the molecular spins, we agree with
them and other authors~\cno{Cuc99,Fer03} that their main effect is
to add an essentially c-number contribution to the bias field on
any given MS. However, we believe that this conclusion was not foregone,
and that our treatment gives a proper justification for neglecting
the additional decoherent effect of these degrees of freedom.

The quasistatic approximation enables us to answer the question posed
at the start, viz., why the LZS formula appears to describe the
swept field experiments so well. We find that this is not because the
transitions are coherent, but because the effective width of the
crossing and the incoherent spin-flip rate vary inversely, leading
to a fortuitous cancellation. It remains an open question to study
the stochastic variation of the bias field, and thus understand this
process even better.

%
%

\acknowledgments
This work was begun with support from the NSF via grant number
DMR-0202165. We are indebted to V. V. Dobrovitski, A. D. Kent,
J.~Villain, and W.~Wernsdorfer for useful discussions and
correspondence.

\appendix
\section{Calculation of Single-Spin Influence Factors for Molecular
Spins}
\label{F_calc}
In this Appendix, we calculate the factor $F_i$ in \eno{infn_up}.
To save writing, we omit the index $i$ henceforth. Let us denote
the influence factor by $\Fu$ or $\Fd$ when the environmenal spin
is up or down respectively. We further abbreviate
\bea
\eps_{\pm} &=& \eps \pm K, \\
\Om_{\pm}^2 &=& \Dta^2 + \eps_{\pm}^2,
\eea
and $t_{12} = t_1 - t_2$ as before.

Let us find $\Fu$ first. As a first approximation, we argue that
because $\Dta$ is much smaller than the typical value of $\eps$ or
$K$, we may neglect it altogether. This yields
\beq
\Fu \apx e^{iKt_{12}}.
\eeq
This approximation is too crude. It implies $|\Fu| = 1$, which is
the maximum possible value it can have. Decoherence arises precisely
from the fact that $|\Fu| < 1$ because $[\hp,\hm] \ne 0$. It is
important to find the departure from unity. With this in mind, let us
write
\beq
\Fu = e^{i\phi} (1 - \eta),  \label{Fpolar}
\eeq
where $\phi$ is a real phase, and $\eta$ is another real quantity that
we have refered to as the {\it mismatch\/}. Our crude calculation
shows that $\eta$ is small and $\phi \apx Kt_{12}$.

Before calculating $\eta$ more carefully, let us relate $\Fd$ to $\Fu$.
Since $\ket\dn = -i \sig_y \ket\up$, we can write
\beq
\Fd = \mel{\up}{\sig_y
                     {e^{i\hp t_1}
                        \sig_y \sig_y
                           e^{-i\hm(t_1 - t_2)}
                             \sig_y \sig_y
                                 e^{-i\hp t_2}} \sig_y} {\up},
\eeq
and since $\sig_y$ anticommutes with $\ham_{\pm}$, this can be
transformed to
\beq
\Fd = \mel{\up}
                     {e^{-i\hp t_1}
                           e^{i\hm(t_1 - t_2)}
                                 e^{i\hp t_2}} {\up},
\eeq
which is the same expression as $\Fu$ with the signs of $t_1$ and $t_2$
reversed. That is,
\beq
\Fd(t_1, t_2) = \Fu(-t_1, -t_2).
\eeq
In particular, we shall see that the mismatch for $\Fd$ is the same as
that for $\Fu$, so that
\beq
\Fd(t_1, t_2) \apx e^{-iK t_{12}} (1-\eta).
\eeq

To find the mismatch more accurately, we write
\beq
\Fu = \tran{\nhat_1}{\nhat_2},
\eeq
where
\beq
\ket{\nhat_a} = e^{i\hm t_a} e^{-i\hp t_a} \ket{\up}, \quad a=1, 2.
      \label{coh1}
\eeq
The notation in this equation exploits the fact that every pure
state of a spin-1/2 system can be written as a spin-coherent-state,
i.e. a state with maximal spin projection along {\it some\/}
direction in space. Thus, the states defined in \eno{coh1}
have maximal spin projections along directions $\nhat_1$ and $\nhat_2$.
These directions remain to be found. Of course, the states also
have phases which also need to be found.

Let us now view \eno{coh1} in terms of two rotations applied to the
state $\ket{\up}$. Since $\Dta \ll \eps$, these rotations are both
about directions very close to $\zhat$. Accordingly, $\nhat_1$
and $\nhat_2$ are also very close to $\zhat$, and we may write
\beq
n_{az} \apx 1 - \tshf n_{a\perp}^2,
\eeq
where $\bn_{a\perp}$ is the component of $\nhat_a$ perpendicular
to $\zhat$. Now, since
\beq
|\tran{\nhat_1}{\nhat_2}|^2 = \hf(1 + \nhat_1\cdot\nhat_2)
\eeq
for spin-1/2 coherent states, we may write
\bea
|\Fu|^2 &\apx& \hf\lf(2 - \hf n_{1\perp}^2 - \hf n_{2\perp}^2
                   +\bn_{1\perp}\cdot\bn_{2\perp} \rt) \\
      &=& 1 - {1\by 4} (\bn_{1\perp} - \bn_{2\perp})^2.
\eea
Taking the square root, and recalling the definition of the
mismatch, we get
\beq
\eta \apx {1\by 8} (\bn_{1\perp} - \bn_{2\perp})^2.
\eeq

The problem is thus to find $\bn_{a\perp}$. We have not been
able to find any simple way to do this except by explicit expansion
and multiplication of the exponentiated operators in \eno{coh1}.
The resulting trigonometric expressions can be made somewhat
easier to handle if we introduce the abbreviations
\beq
\tta_{1\pm} = \Om_{\pm} t_1/2, \quad
  \tta_{2\pm} = \Om_{\pm} t_2/2.
\eeq
With these, we may write
\beq
e^{-i\hp t_2} = c_0 + \bc\cdot{\vec\sig},
\eeq
where
\beq
c_0 = \cos \tta_{2+}, \quad
\bc = -{i\by\Opl}\sin\tta_{2+} (\Dta, 0, \epl).
\eeq
Similarly,
\beq
e^{i\hm t_2} = b_0 +\bb\cdot{\vec\sig},
\eeq
with
\beq
b_0 = \cos \tta_{2-}, \quad
\bb = {i\by\Omi}\sin\tta_{2-} (\Dta, 0, \emi).
\eeq
Then,
\beq
\ket{\nhat_2} = (b_0 +\bb\cdot{\vec\sig})
                  (c_0 + \bc\cdot{\vec\sig}) \ket{\up}
              = A_{2\up}\ket{\up} + A_{2\dn} \ket{\dn},
                             \label{Aform}
\eeq
with
\bea
A_{2\up} &=& (b_0 + b_z)(c_0 + c_z) + b_x c_x, \\
A_{2\dn} &=& (b_0 - b_z)c_x + b_x (c_0 + c_z).
\eea
In terms of these quantities, we have
\bea
n_{2+} &=& n_{2x} + in_{2y} \nnu \\
         &=& \mel{\nhat_2}{\sig_+}{\nhat_2} \nnu\\
         &=& 2 A^*_{2\up} A_{1\dn}.
\eea
By comparing and real and imaginary parts of both sides, we
obtain $n_{2x}$ and $n_{2y}$. We now note that the quantities
$A_{2\up}$ and $A_{2\dn}$ consist of various terms oscillating
at the sums and differences of the frequencies $\Om_{\pm}/2$.
Since $\Dta \ll \Om_{\pm}$ for all but very distant (and therefore
very weakly coupled) MS's , we may expand the amplitudes of
these oscillatory factors in powers of $\Dta$. Using the results
\beq
{\epl\by\Opl} \simeq 1 - {\Dta^2 \by 2\Opl^2},
\eeq
etc., we obtain
\bea
c_0 + c_z &=& e^{-i\tta_{2+}} + O(\Dta)^2, \\
b_0 \pm b_z &=&  e^{\pm i\tta_{2-}} + O(\Dta)^2.
\eea
Therefore,
\bea
A_{2\up} &=& e^{i(\tta_{2-} - \tta_{2+})} + O(\Dta)^2, \\
       \label{A2up}
A_{2\dn} &=&
    -i {\Dta\by \Opl} \sin \tta_{2+} e^{-i \tta_{2-}}
    +i {\Dta\by \Omi} \sin \tta_{2-} e^{-i\tta_{2+}},
\eea
and
\beq
A^*_{2\up} A_{2\dn} = 
    -i \lf[{\Dta\by \Opl} \sin\tta_{2+}
                      e^{-i(2\tta_{2-} - \tta_{2+})}
    - {\Dta\by \Omi} \sin\tta_{2-} e^{-i\tta_{2-}}\rt].
\eeq
From this expression, we can get $n_{2x}$ and $n_{2y}$ by taking
and real and imaginary parts.
\bea
n_{2x} &=& - 2{\Dta\by \Opl}\sin \tta_{2+}
                   \sin(2\tta_{2-} - \tta_{2+})
           + 2{\Dta\by\Omi} \sin^2 \tta_{2-} \\
       &=& -{\Dta\by \Opl} [\cos 2(\tta_{2+} - \tta_{2-})
                             - \cos 2\tta_{2-}]
           +{\Dta\by \Omi} [1- \cos 2\tta_{2-}] \\
       &=& {\Dta\by \Omi} - {2\Dta K\by \Opl\Omi} \cos2\tta_{2-}
                  -{\Dta\by \Opl} \cos 2(\tta_{2+} - \tta_{2-}),
\eea
where in the last line we have used the result
\beq
{1\by \Opl} - {1\by \Omi} \simeq -{2K\by \Opl\Omi}.
\eeq
In the same way, we have
\bea
n_{2y} &=& - 2{\Dta\by \Opl}\sin \tta_{2+}
                   \cos(2\tta_{2-} - \tta_{2+})
           + 2{\Dta\by\Omi} \sin\tta_{2-} \cos\tta_{2-} \\
       &=& -{\Dta\by \Opl} [\sin 2(\tta_{2+} - \tta_{2-})
                             + \sin 2\tta_{2-}]
           +{\Dta\by \Omi} \sin 2\tta_{2-}  \\
       &=& {2\Dta K\by \Opl\Omi} \sin2\tta_{2-}
                  -{\Dta\by \Opl} \sin 2(\tta_{2+} - \tta_{2-}).
\eea
For $n_{1x}$ and $n_{1y}$, we simply change the suffix 2 {}in
$\tta_{2\pm}$ from 2 to 1. We then have
\bea
\!\!\!n_{1x} - n_{2x} &=& {2\Dta K\by \Opl\Omi}
                        [\cos2\tta_{2-} - \cos2\tta_{1-}]
                  + {\Dta\by \Opl}
                        [\cos 2(\tta_{2+} - \tta_{2-})
                            - \cos 2(\tta_{1+} - \tta_{1-})],\ \ \\
\!\!\!n_{1y} - n_{2y} &=&  -{2\Dta K\by \Opl\Omi}
                        [\sin2\tta_{2-} - \sin2\tta_{1-}]
                    + {\Dta\by \Opl}
                        [\sin 2(\tta_{2+} - \tta_{2-})
                            - \sin 2(\tta_{1+} - \tta_{1-})].
\eea
We can simplify these expressions by first noting that
\beq
\tta_{a+} - \tta_{a-}
   \simeq Kt_a \lf(1 + O(\Dta/\eps_{\pm})^2 \rt), \quad (a = 1, 2)
           \label{tpm}
\eeq
and then defining sum and difference variables
\beq
\bar t = \tshf(t_1 + t_2), \quad t_{12} = t_1 - t_2,
\eeq
in terms of which we have identities like
\bea
\cos2\tta_{2-} - \cos2\tta_{1-}
    &=& 2\sin\Omi\bar t \sin \tshf\Omi t_{12}, \\
\sin2\tta_{2-} - \sin2\tta_{1-}
    &=& -2\cos\Omi\bar t \sin \tshf\Omi t_{12},
\eea
etc. Putting all these together, we obtain
\bea
n_{1x} - n_{2x} &=&
     {4\Dta K \by \Opl\Omi}
           \sin\Omi\bar t \sin \tshf\Omi t_{12}
    +{2\Dta\by\Opl}
           \sin 2K\bar t \sin K t_{12}, \\
n_{1y} - n_{2y} &=&
     {4\Dta K \by \Opl\Omi}
           \cos\Omi\bar t \sin \tshf\Omi t_{12}
    -{2\Dta\by\Opl}
           \cos 2K\bar t \sin K t_{12}.
\eea
Squaring and adding these two expressions, we obtain the mismatch as
\beq
\eta = 2 \lf( {\Dta K \by \Opl\Omi} \rt)^2
           \sin^2 \tshf\Omi t_{12}
    + \hf \lf({\Dta\by\Opl}\rt)^2 \sin^2 K t_{12}
    - 2 {\Dta^2 K \by \Opl^2 \Omi}
             \cos\Opl \bar t \sin \tshf\Omi t_{12}
                                \sin K t_{12}.
     \label{mis}
\eeq
We can also write this in a manifestly positive form:
\beq
\eta = {\Dta^2 \by 2\Opl^2}
       \lf[ 2 {K\by \Omi} \sin {\Omi t_{12} \by 2}
                                       - \sin K t_{12} \rt]^2
      + 4 {\Dta^2 K \by \Opl^2 \Omi}
             \sin^2 \tshf\Opl \bar t
               \sin \tshf\Omi t_{12}
                  \sin K t_{12}.
           \label{mis2}
\eeq
Since this expression is unchanged when the signs of both
$t_1$ and $t_2$ are reversed, we have now proven our claim that it
is also the mismatch for $\Fd$. \Etwo{mis}{mis2} are valid for
$t_{1,2} \ll E_{dm}/\Dta^2$ on account of \eno{tpm}.

To find the phase $\phi$ in \eno{Fpolar}, we note that from
\etwo{A2up}{tpm} that
\beq
A_{2\up} = e^{-iKt_2 + O(\Dta)^2} + O(\Dta)^2,
\eeq
and likewise for $A_{1\up}$. Now, since $A_{a\dn} = O(\Dta)$,
\bea
\tran{\nhat_1}{\nhat_2}
    &=& A^*_{1\up} A_{2\up} + A^*_{1\dn} A_{2\dn} \\
    &=& A^*_{1\up} A_{2\up} + O(\Dta)^2 \\
    &=& e^{iK t_{12}} + O(\Dta)^2.
\eea
Thus, the dominant term in $\phi$ is just the
zeroth order one, i.e., $\phi \apx K t_{12}$, and we have,
\beq
\Fu(t_1, t_2) \apx e^{iK t_{12}} (1-\eta),
\eeq
with $\eta$ given by \eno{mis}. As a check, note that $\Fu$ correctly
vanishes when $t_{12} = 0$.

The calculation above assumes that $K \sim \eps \gg \Dta$. We shall
see in Appendix \ref{eta_estimate} that distant spins for which
$K \sim \Dta$ play an important role in determining the net influence
factor. It is therefore desirable to find $\eta$ when $K$ is small.
This can be done by evaluating \eno{infn_up} for $\Fu$ by a
standard perturbation expansion in $K$. The result is
\beq
\eta = {K^2 \eps^2 \Dta^2 \by 2 \Om^4} t_{12}^2
       - {K^2 \eps^2 \Dta^2 \by \Om^5} t_{12} \sin \Om t_{12}
       + 2 {K^2 \Dta^2 \by \Om^4} \sin^2 \tshf\Om t_{12}
       - {K^2 \Dta^4 \by 2 \Om^6} \sin^2 \Om t_{12}.
\eeq
Here, $\Om = (\eps^2 + \Dta^2)^{1/2}$. The last term is smaller than
the first three by order $(\Dta/E_{dm})^2$. The remaining three terms
are qualitatively very similar to what we get from \eno{mis} when
$K \to 0$.
\section{Estimate of Multi-Spin Influence Factor for Molecular
Spin Environment}
\label{eta_estimate}

In this appendix, we will estimate the total influence factor
$F = \prod_i F_i$ for $|t_{12}| \gg E_{dm}^{-1}$, using a mix of
analytic and numerical approaches.

\subsection{Preliminary Analytic Estimate}

Since $\eta_i \ll 1$, the total influence factor is given by
\beq
F \simeq e^{i \sum_i K_i s_i} e^{-\sum_i \eta_i},
\eeq
so $|F| \simeq \exp(-\sum_i \eta_i)$. We therefore focus on the
sum $\sum_i \eta_i$. Let us divide it into three parts, $S_1$, $S_2$,
and $S_3$, corresponding to the three terms in $\eta_i$ in \eno{mis}.

The first sum is
\beq
S_1 = 2 \sum_i \Dta^2 \lf(\frac{K_i}{\Om_{i+}}\rt)^2
              \lf(\frac{\sin(\Om_{i-}t_{12}/2)}{\Om_{i-}}\rt)^2.
    \label{S1_1}
\eeq
For large $|t_{12}|$ (but with $|t_{12}| \ll \Dta^{-1}$), we may
replace the factor
$ \Oimi^{-2} \sin^2(\Om_{i-}t_{12}/2)$
by a term proportional to $\dta(\Oimi) |t_{12}|$ as in textbook
derivations of Fermi's golden rule. The replacement must be done with
care, however. The physical point is that for large $|t_{12}|$, the only
sites that contribute significantly to $S_1$ are those for which
$\Oimi$ is very small. By taking the distribution as $\dta(\Oimi)$,
we get a vanishing answer for $S_1$ since $\Oimi\geq\Dta$,
and so the argument of the $\dta$-function is never satisfied. The
correct result which preserves the integral with respect to $\Oimi$ is
\beq
\lf[\frac{\sin(\Om_{i-}t_{12}/2)}{\Om_{i-}}\rt]^2
    \apx \frac{\pi}{4}\delta(\Oimi - \Dta) |t_{12}|.
     \label{S1_dta}
\eeq
In this equation, the weight of the delta-function is
$\pi/4$ instead of $\pi/2$ since on the left we only integrate
over positive values of $\Oimi$, but on the right we wish to interpret
the delta-function in the standard way, that is, as a distribution to
be integrated over all $\Oimi$. Using \eno{S1_dta} in \eno{S1_1}
yields
\beq
S_1 = {\pi\over 2} \sum_i \Dta^2 {K_i^2 \by \Oipl^2}
                             \dta(\Oimi - \Dta) |t_{12}|.
\eeq
To further simplify this result, we note that $\Oimi = \Dta$
implies $\eimi = 0$, $\eipl = 2K_i$, and $\Oipl^2 = \Dta^2 + 4K_i^2$.
Therefore,
\beq
\dta(\Oimi - \Dta) = {\Oimi \by |\eps_i|} \dta(\eps_i - K_i)
                   = {\Dta \by |K_i|} \dta(\eps_i - K_i),
\eeq
and
\beq
S_1 = \hf \pi \Dta^3 |t_{12}|
          \sum_i {|K_i| \by 4K_i^2 + \Dta^2} \, \dta(\eps_i - K_i).
\eeq
We now average over the bias distribution (\ref{dist_eps}). This turns
$\dta(\eps_i - K_i)$ into $f(K_i)$. It then remains to do the sum
over the sites. Because the summand is slowly varying, we may replace
the sum by an integral. This integral may in turn be performed by
introducing the density of couplings $g(K)$, defined so that $g(K)dK$
is the number of sites for which $K_i$ lies between $K$ and $K + dK$.
In this way we get
\beq
S_1 = \hf \pi \Dta^3 |t_{12}|
         \int_{|K| > c\Dta} dK\, {|K| \by 4K^2 + \Dta^2}\, f(K) g(K).
\eeq
We have cut off the $K$ integration so as to exclude
very distant spins for which the coupling is weaker than $c\Dta$,
where $c$ is some constant of order unity. The reason is that for
such spins the mismatch will be essentially zero, since they
are insensitive to the orientation of the central spin.

We show in Appendix \ref{dens_coup} that
\beq
g(K) = {16 \pi \by 9 \sqrt{3}} {E_{dm} \by K^2}.
     \label{g_coup}
\eeq
Note that couplings $+K$ and $-K$ are equally likely. Using this
result, we obtain
\beq
S_1 = {8\pi \by 9}\sqrt{2\pi \by 3} {\Dta^3 E_{dm} \by E_b} |t_{12}|
          \int_{c\Dta}^{\infty} dK
               {1 \by K(4K^2 + \Dta^2)} e^{-K^2/ 2E_b^2}.
\eeq
The integral is dominated by small values of $K$ close to $\Dta$, so
it may be evaluated by parts. Doing so, and setting
$e^{-\Dta^2 /2 E_b^2} \simeq 1$, we obtain
\beq
S_1 = {4\pi \by 9} \sqrt{2\pi\by 3}
         \ln\lf(1 + {1\by 4 c^2} \rt)
           {\Dta E_{dm} \by E_b} |t_{12}|.
    \label{ans_S1}
\eeq
Note that the Gaussian form of $f(\eps)$ is not essential to the
form of the answer.

The second term in \eno{mis} leads to the sum
\beq
S_2 = \hf \sum_i \frac{\Dta^2}{\Om_{i+}^2}\sin^{2}(K_i t_{12}). 
\eeq
Since $\sin^{2}(K_i t_{12})$ is bounded by 1 the dominant contribution
will come from sites on which $\Om_{i+}^2\apx O(\Dta^2)$. Averaging
over the bias field distribution gives
\beq
S_2 = \hf \sum_i \inmipi {\Dta^2 \by \Dta^2 + (\eps + K_i)^2}
                     f(\eps) \sin^2(K_i t_{12})\, d\eps.
\eeq
Because $\Dta \ll E_{dm}$, the integral is very sharply peaked
at $\eps = - K_i$. We may therefore replace $f(\eps)$ by the
constant $f(-K_i)$. The integral is then elementary, and we obtain
\beq
S_2 = \hf \pi \Dta \sum_i f(-K_i) \sin^2(K_i t_{12}).
\eeq
The sum is now evaluated as before, by converting to an integral
over $K$. Using the result (\ref{g_coup}) for $g(K)$, we obtain
\beq
S_2  = {8 \pi^2 \by 9\sqrt{3}} \Dta E_{dm} \int_{|K| > c\Dta} dK
          f(-K) {\sin^2 (K t_{12}) \by K^2}.
\eeq
This time the integrand is sufficiently convergent near $K =0$,
so the limit $c\Dta$ can be replaced by 0. 
The factor $K^{-2} \sin^2(K t_{12})$ behaves like
$\pi |t_{12}| \dta(K)$ for large $|t_{12}|$. We may therefore replace
$f(-K)$ by $f(0) = (2\pi E_b^2)^{-1/2}$, after which the integral is
trivial, and yields
\beq
S_2 = {4\pi^2\by 9}\sqrt{2\pi \by 3} {\Dta E_{dm} \by E_b} |t_{12}|.
   \label{ans_S2}
\eeq
This result is also valid only for $|t_{12}| \ll \Dta^{-1}$.

The last sum, $S_3$, from the third term in \eno{mis}, is given by
\beq
S_3 = 2 \sum_i\frac{\Dta^2}{\Om_{i+}^2} \frac{K_i}{\Om_{i-}}
            \sin(\Om_{i-}t_{12}/2)
              \sin(K_i t_{12}) \cos(\Oipl \bar t).
\eeq
As $|t_{12}|$ increases, the term $\Oimi^{-1} \sin(\Om_{i-}t_{12}/2)$
behaves like a $\delta$-function of $\Oimi$. By the same reasoning
as for $S_1$, we find that the correct replacement is
\beq
{\sin(\Om_{i-}t_{12}/2) \by \Oimi}
   = {\pi \by 2} \delta(\Oimi -\Dta)\, {\rm sgn}(t_{12}).
     \label{S3_dta}
\eeq

Further writing
$\Om_{i\pm}$ in terms of $\Dta$, $\eps_i$, and $K_i$, we obtain
\beq
S_3 = \pi \sum_i {\Dta^3 \by 4K_i^2 + \Dta^2}
            \sin(|K_i t_{12}|)
               \cos(\sqrt{4K_i^2 + \Dta^2} \bar t)
                 \, \dta(\eps_i - K_i),
\eeq
where we have incorporated the ${\rm sgn}(t_{12})$ and ${\rm sgn}\,K_i$
factors by taking an absolute value of the argument in
$\sin(|K_i t_{12}|)$. The next step is to average over the bias
distribution and integrate over the sites. As in the case of $S_1$,
we exclude distant spins, and obtain,
\beq
S_3 = {16\by9} \sqrt{2\pi^3 \by 3} {\Dta^3 E_{dm} \by E_b}
       \int_{c\Dta}^{\infty} dK {1 \by K^2 (4K^2 + \Dta^2)}
              \sin(K |t_{12}|)
                \cos(\sqrt{4K^2 + \Dta^2} \bar t)
                 e^{-K^2/2E_b^2}.
\eeq
This integral is also dominated by the lower limit, but the answer
is different depending on $\bar t$. If $\bar t \ll \Dta^{-1}$,
we may argue that for $K \sim \Dta$,
and for $E_{dm}^{-1} \ll |t_{12}| \ll \Dta^{-1}$,
$\sin(K|t_{12}|) \apx K|t_{12}|$, and
$\cos (\sqrt{4K^2 + \Dta^2} \bar t) \apx 1$. The
resulting integral is identical to that which appeared in $S_1$.
Hence, we have
\beq
S_3 = {8\pi \by 9} \sqrt{2\pi\by 3}
         \ln\lf(1 + {1\by 4 c^2} \rt)
           {\Dta E_{dm} \by E_b} |t_{12}|,
                    \quad (\bar t \ll \Dta^{-1}).
    \label{ans_S3}
\eeq
If on the other hand $\bar t \gtwid \Dta^{-1}$, the oscillations in
the $\cos (\sqrt{4K^2 + \Dta^2} \bar t)$ factor reduce $S_3$
significantly. The precise from is unimportant, and it suffices
to put
\beq
S_3 \simeq 0, \quad (\bar t \gtwid \Dta^{-1}).
\eeq

For short times $\bar t \ll \Dta^{-1}$ (which automatically implies
$|t_{12}| \ll \Dta^{-1}$), all three sums $S_i$, have the same
behaviour. Adding them together, we obtain,
\beq
\sum_i \eta_i
   = S_1 + S_2 - S_3
   = {4\pi \by 9} \sqrt{2\pi\by 3}
       \lf[\pi - \ln\lf(1 + {1\by 4 c^2} \rt) \rt]
           {\Dta E_{dm} \by E_b} |t_{12}|
      \label{eta_lin}
\eeq
On the other hand, for longer $\bar t$, but still obeying
$\Dta|t_{12}| \ll 1$, $S_3$ may be neglected, and
\beq
\sum_i \eta_i
   = {4\pi \by 9} \sqrt{2\pi\by 3}
       \lf[\pi + \ln\lf(1 + {1\by 4 c^2} \rt) \rt]
           {\Dta E_{dm} \by E_b} |t_{12}|
      \label{eta_lin2}
\eeq

\subsection{Improved estimate incorporating numerics}
\label{eta_num}

Since the analytical estimate given above entails several
approximations, we have also evaluated $|F|$ numerically.
We take the contribution $F_i$ from the $i$th MS to be given by
$e^{i s_i K t_{12}} (1-\eta_i)$, with $\eta_i$ given by \eno{mis}.
The different factors $\eta_i$ are found, and the factors
$(1-\eta_i)$ multiplied to obtain $|F|$. This calculation is
valid for a much larger range of times, $t_{1,2} \sim E_{dm}/\Dta^2$,
since the expression (\ref{mis}) then holds.

In more detail, our algorithm is as follows. We first create a
set of sites on a nearly cubic lattice. That is, each site is offset
from a perfect cubic lattice by a small random amount equal to
$0.01$ times the lattice constant in each of
the three cartesian directions. (The reason for adding the offsets
was to avoid exact cancellations of the dipole field from an
aligned shell of nearest neighbor spins. We do not believe that this
step is essential, but it does not invalidate the calculation either.)
We next place a spin on each site with the orientation $s_i$
randomly chosen to be $\pm 1$ with equal probability, and choose
a particular value of $t_{12}$. Next, at each
site we select an energy bias $\eps_i$ by sampling a normal
distribution with mean zero and standard deviation $E_b = 1000$
in units such that $\Dta = 1$. (In \Fe8 the ratio $E_b/\Dta$ is
$\sim 10^6$. Taking such a large ratio in the numerics makes each
individual mismatch prohibitively small, making it very hard to see
departures from unity in $\prod (1-\eta_i)$. The physically important
point is to ensure $E_b/\Dta \gg 1$, which we do.) The dipole field
$K_i$ at each site due to the central spin is computed using
\eno{def_Ki2} with $E_{dm}$ also equal to $1000\Dta$. That is, we do
not take $E_b$ and $E_{dm}$ to be different. With these values of
$K_i$ and $\eps_i$, we can then find (all energies are computed in
units of $\Dta$)
\beq
\eps_{i\pm} = \eps_i \pm K_i,\quad
{\Om_{i\pm}}^2 = {\eps_{i\pm}}^2+{\Dta}^2.
\eeq
It is now possible to calculate the expression (\ref{mis}) for
$\eta_i$ for any $\bar t$ and $t_{12}$. The dependence on two time
variables is inconvenient, however, so instead we calculate the
lower and upper bounds  with respect to $\bar t$, which
are given by 
\bea
\eta_{i\,{\rm min}}
   &=& 2\lf(\frac{\Dta}{\Om_+}\rt)^2
       \lf[\lf(\frac{K_i}{\Om_{i-}}\rt)
                   \sin \lf(\frac{\Om_{i-}t_{12}}{2}\rt)
              - \frac{1}{2}\sin(K_i t_{12}) \rt]^2, \\
                        \label{eq:eta_min}
\eta_{i\,{\rm max}}
   &=& 2\lf(\frac{\Dta}{\Om_+}\rt)^2
       \lf[\lf(\frac{K_i}{\Om_{i-}}\rt)
                   \sin \lf(\frac{\Om_{i-}t_{12}}{2}\rt)
              + \frac{1}{2}\sin(K_i t_{12}) \rt]^2.
                        \label{eq:eta_max}
\eea
Since $|F| = \prod_i |1-\eta_i|$, we have 
\beq
\prod_i(1-\eta_{i\,{\rm max}})
     \le  |F| \le \prod_i (1-\eta_{i\,{\rm min}}).
\eeq
These bounds, $|F|_{\rm min}$ and $|F|_{\rm max}$, are now found
using the computed maxima and minima for $\eta_i$. At the same time,
we also compute the sums $S_1$, $S_2$, and
\beq
S'_3 = 2 \sum_i\frac{\Dta^2}{\Om_{i+}^2} \frac{K_i}{\Om_{i-}}
            \sin(\Om_{i-}t_{12}/2)
              \sin(K_i t_{12}),
\eeq
which differs from $S_3$ in that the factor $\cos(\Oipl \bar t)$
is lacking in the summand. As argued above, $S'_3 \simeq S_3$
when $\Dta \bar t \ll 1$. In this time range, therefore, we
expect $|F| \simeq |F|_{\rm max}$. For longer $\bar t$ on the
other hand, $|F| \simeq \exp(-(S_1 + S_2))$. We nevertheless
shall find it useful to continue to calculate $S'_3$. 

We then recompute the $S$'s, and the bounds for $|F|$ using a different
set of biases, $\eps_i$. All told, we do this for about $10^5$ bias
configurations, in order to generate averages for the $S$'s, and
$|F|_{\rm min}$ and $|F|_{\rm max}$. We also calculate the
variances in these quantities at this stage. The entire calculation
is then repeated for different $t_{12}$.

The lower and upper bounds of $|F|$ , $|F|_{\rm min}$ and
$|F|_{\rm max}$, are plotted as a function of $t_{12}$
in Figs.~\ref{Dta0p1_8000.eps}--\ref{Dta0p5_27000.eps}.
(We do not show the data for $N=8000$ and $\Dta/E_{dm}$ values
of $0.003$ or $0.005$.)

We see that $|F|_{\rm min}$ does die as an exponential,
i.e., we can fit it to a form $e^{-a|\Dta t_{12}|}$ very well. On the
other hand, $|F|_{\rm max}$ dies like $e^{-b |\Dta t_{12}|^2}$, which is
rather different. We show the best fit values of $a$ and $b$ for
three different $\Dta/E_{dm}$ in \Tno{tab1}. As can be seen,
$a$ and $b$ are reasonably independent of this ratio.
 
The result $|F|_{\rm max} \sim e^{-b (\Dta t_{12})^2}$ is rather
surprising, since
\bea
|F|_{\rm min} &\simeq& e^{-(S_1 + S_2 + S'_3)}, \\
|F|_{\rm max} &\simeq& e^{-(S_1 + S_2 - S'_3)},
\eea
and we showed that all three sums vary linearly with
$|t_{12}|$. We can understand the $t_{12}^2$ behaviour from our
numerics. We first note that we do indeed find an excellent linear
$|t_{12}|$ variation for the the individual $S$'s. We therefore write
$S_i = \zeta_i \Dta |t_{12}$, and determine $\zeta_i$ from our data.
These values are shown in \Tno{tab2}. Examining the table,
we see that (a) $\zeta_1 \simeq \zeta_2$, and (b) there is a
nearly total cancellation in $\zeta_1 + \zeta_2 - \zeta'_3$. i.e.,
$\zeta'_3 \simeq \zeta_1 + \zeta_2$.

We can use the numerical results to improve our analytical estimate
as follows. We assume that the above mentioned cancellation is perfect.
In other words, the unknown parameter $c$ in \eno{eta_lin} is such
that
\beq
\ln \lf(1 + {1\by 4c^2} \rt) = \pi.
                        \label{fix_c}
\eeq
This implies that
\bea
\sum_i \eta_{i\,{\rm max}}
   = S_1 + S_2 + S'_3
   &=& {4\pi \by 9} \sqrt{2\pi\by 3}
       \lf[\pi + 3 \ln\lf(1 + {1\by 4 c^2} \rt) \rt]
           {\Dta E_{dm} \by E_b} |t_{12}| \\
   &=& {16\pi^2 \by 9} \sqrt{2\pi\by 3}
           {\Dta E_{dm} \by E_b} |t_{12}|.
   \label{sum_S}
\eea
With $E_{dm} = E_b$, this equals $25.4 \Dta t_{12}$. Our numerical
fits to $|F|_{\rm min}$ yield $a \simeq 24$, which is quite close.
This gives us confidence in fixing $c$ as per
\eno{fix_c}~\cno{quad_Fmax}.

We can now estimate $\sum_i \eta_i$ for long $\bar t$ (but
$\ll E_{dm}^2/\Dta$). Using \eno{fix_c}, we have
\beq
\sum_i \eta_i \simeq S_1 + S_2
   = {8\pi^2 \by 9} \sqrt{2\pi\by 3}
           {\Dta E_{dm} \by E_b} |t_{12}|
      \label{eta_lin3}
\eeq
Since we do not know $E_{dm}/E_b$ precisely, however, we limit ourselves
to stating that
\beq
\sum_i \eta_i \simeq \gam_m \Dta |t_{12}|,
  \label{net_mis}
\eeq
where $\gam_m$ is a constant of order unity.

\section{Density of dipole coupling strengths}
\label{dens_coup}

The density of dipole couplings, $g(K)$, introduced in \eno{g_coup},
is given by
\beq
g(K) = \sum_i \dta(K-K_i),
\eeq
with
\beq
K_i = {2E_{dm} a^3 \by r_i^3} (1-3\cos^2\tta_i). 
\eeq
We evaluate the sum over lattice sites assuming that the spins are
uniformly distributed with a density $a^{-3}$. Except when
$K \simeq E_{dm}$, corresponding to nearest or next-nearest
neighbour sites, we may replace the sum by an integral, obtaining
\beq
g(K) = {2\pi \by a^3}\inzpi dr\, r^2
         \int_{-1}^{1} du \, 
            \delta \lf[K - {2E_{dm}a^3 \by r^3} (1-3u^2) \rt],
\eeq
where $u = \cos\tta$. Performing the $r$ integral, we get
\beq
g(K) = {4\pi E_{dm} \by 3 K^2}
         \int_{-1}^{1} du \, 
             |1 -3 u^2| \,\Tta\lf({1-3u^2 \by K} \rt),
\eeq
where $\Tta(\cdot)$ is the Heavyside step function; equal to 1 when its
argument is positive, and zero otherwise. The integral on $u$ is
best done separately for positive and negative $K$. When $K > 0$,
we have
\beq
g(K) = {8\pi E_{dm} \by 3 K^2}
         \int_{0}^{3^{-1/2}} du \, (1 -3 u^2)
     = {16 \pi \by 9 \sqrt{3}} {E_{dm} \by K^2}.
\eeq
Likewise, when $K < 0$, we have
\beq
g(K) = {8\pi E_{dm} \by 3 K^2}
         \int_{3^{-1/2}}^1 du \, (3 u^2 - 1)
     = {16 \pi \by 9 \sqrt{3}} {E_{dm} \by K^2},
\eeq
the same expression as for $K > 0$. This is \eno{g_coup}.

\newpage
\begin{table}
\caption{\label{tab1} Best fit values of the parameters $a$ and
$b$.}
\begin{ruledtabular}
\begin{tabular}{c c c c c}
    & \muc{2}{c}{$a$} & \muc{2}{c}{$b$} \\
\mbox{$\Dta/E_{dm}$}
    & \mbox{$N = 8000$} & \mbox{$N = 27000$}
    & \mbox{$N = 8000$} & \mbox{$N = 27000$} \\
\hline
   0.001 & 23.7 & 24.4 & 2.09 & 2.80 \\
   0.003 & 24.0 & 24.2 & 2.52 & 2.65 \\
   0.005 & 23.1 & 23.3 & 2.64 & 2.77 \\
\end{tabular}
\end{ruledtabular}
\end{table}

\begin{table}
\caption{\label{tab2} Numerically calculated values of the coefficients
$\zeta_1$, $\zeta_2$, and $\zeta'_3$ in the sums $S_1$, $S_2$,
and $S'_3$.}
\begin{ruledtabular}
\begin{tabular}{c c c c c c c}
    & \muc{3}{c}{$N= 8000$} &
               \muc{3}{c}{$N = 27000$} \\
\mbox{$\Dta/E_{dm}$}
    & \mbox{$\zeta_1$} & \mbox{$\zeta_2$} & \mbox{$\zeta'_3$}
    & \mbox{$\zeta_1$} & \mbox{$\zeta_2$} & \mbox{$\zeta'_3$} \\
\hline
   0.001 & 5.44 & 5.41 & 10.42    & 6.06 & 6.06 & 11.82 \\
   0.003 & 6.08 & 6.08 & 11.90    & 6.12 & 6.11 & 11.77 \\
   0.005 & 5.98 & 5.99 & 11.78    & 6.03 & 6.11 & 11.01 \\
\end{tabular}
\end{ruledtabular}
\end{table}

\newpage
\begin{figure}
\includegraphics{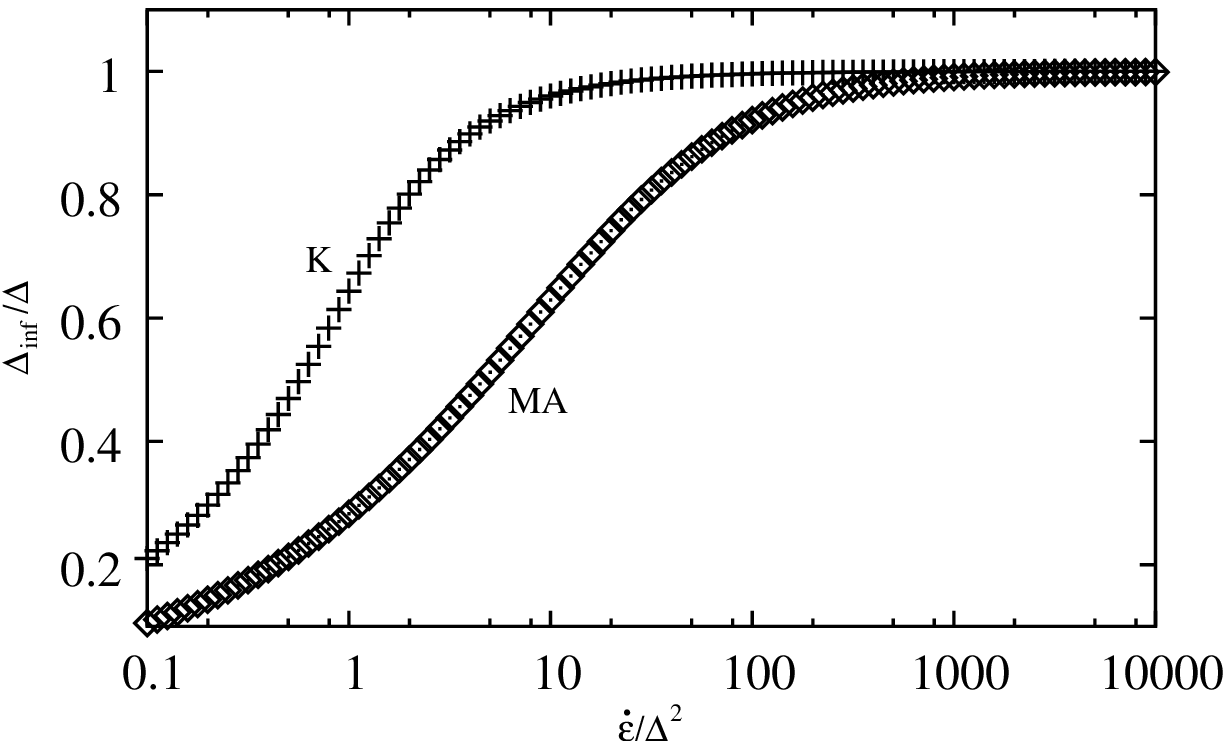}
\caption{\label{Dta_inf.eps} Inferred values of tunnel splitting
as a function of the rate at which the applied field is swept,
assuming that the spin flip probability is given by the
Landau-Zener-Stuckelberg formula, \eno{p_LZS}. The curves marked
K and MA are obtained when the true flip probability is taken to obey
Kayanuma's formula, \eno{pf_qs}, and the macroscopically averaged
formula obtained by integrating \eno{pf_MA}. For the latter, we
took $4\al E_{dm}/W = 40$. This figure should be compared with
Fig. 7 of Ref.~\cno{ww2}.}
\end{figure}
%
\begin{figure}
\includegraphics{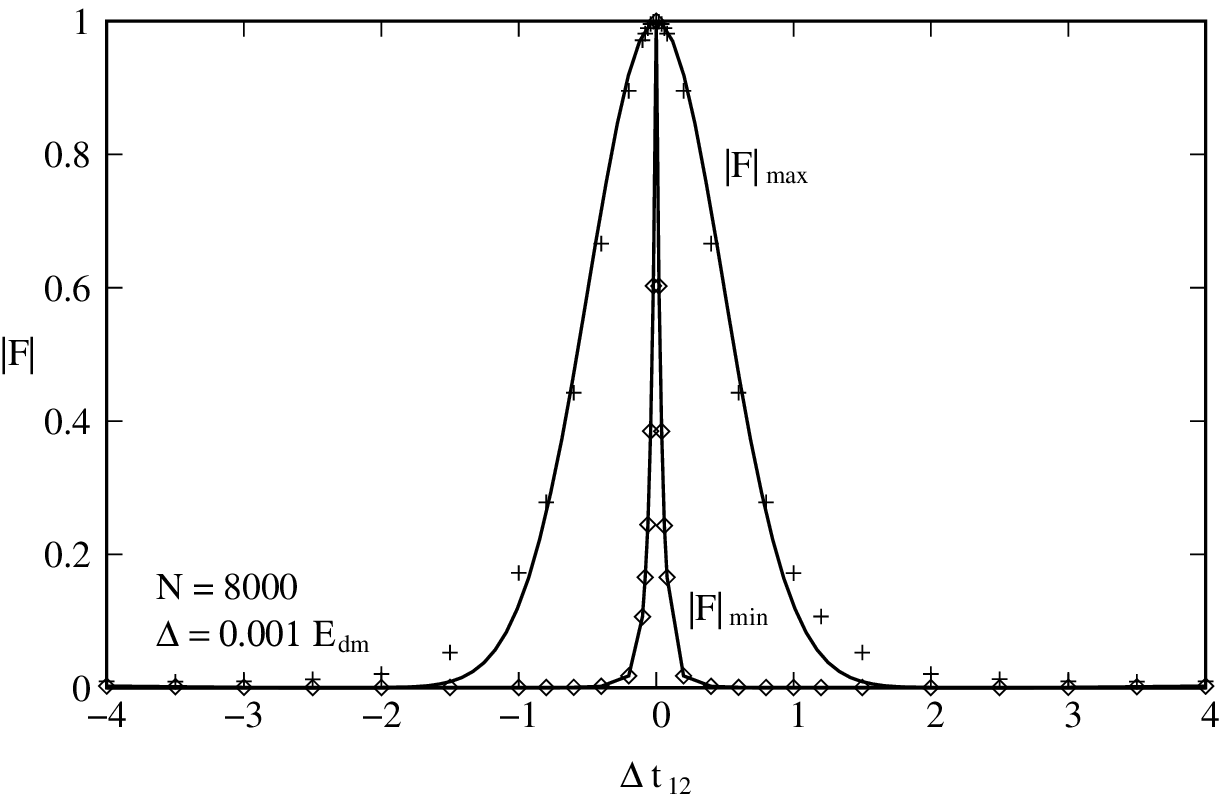}
\caption{\label{Dta0p1_8000.eps} Numerically computed lower and upper
bounds $|F|_{\rm min}$ and $|F|_{\rm max}$, plotted vs. $\Dta t_{12}$,
for a central spin in a lattice of 8000 spins. We have chosen
$\Dta = 0.001 E_{dm}$. The curves are best fits to
$e^{-a|\Dta t_{12}|}$ for $|F|_{\rm min}$, and $e^{-b |\Dta t_{12}|^2}$
for $|F|_{\rm max}$.} 
\end{figure}
\begin{figure}
\includegraphics{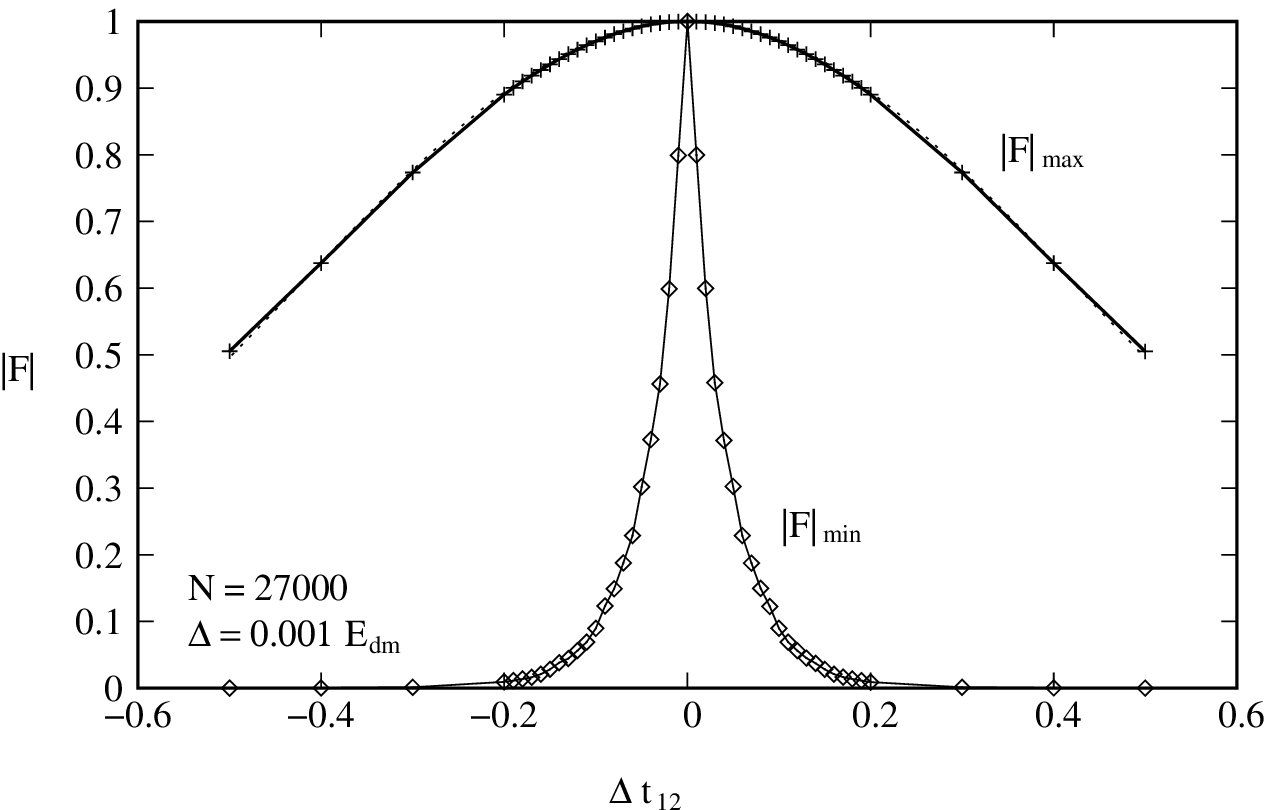}
\caption{\label{Dta0p1_27000.eps}
Same as \fno{Dta0p1_8000.eps} for a lattice of 27000 spins.}
\end{figure}
\begin{figure}
\includegraphics{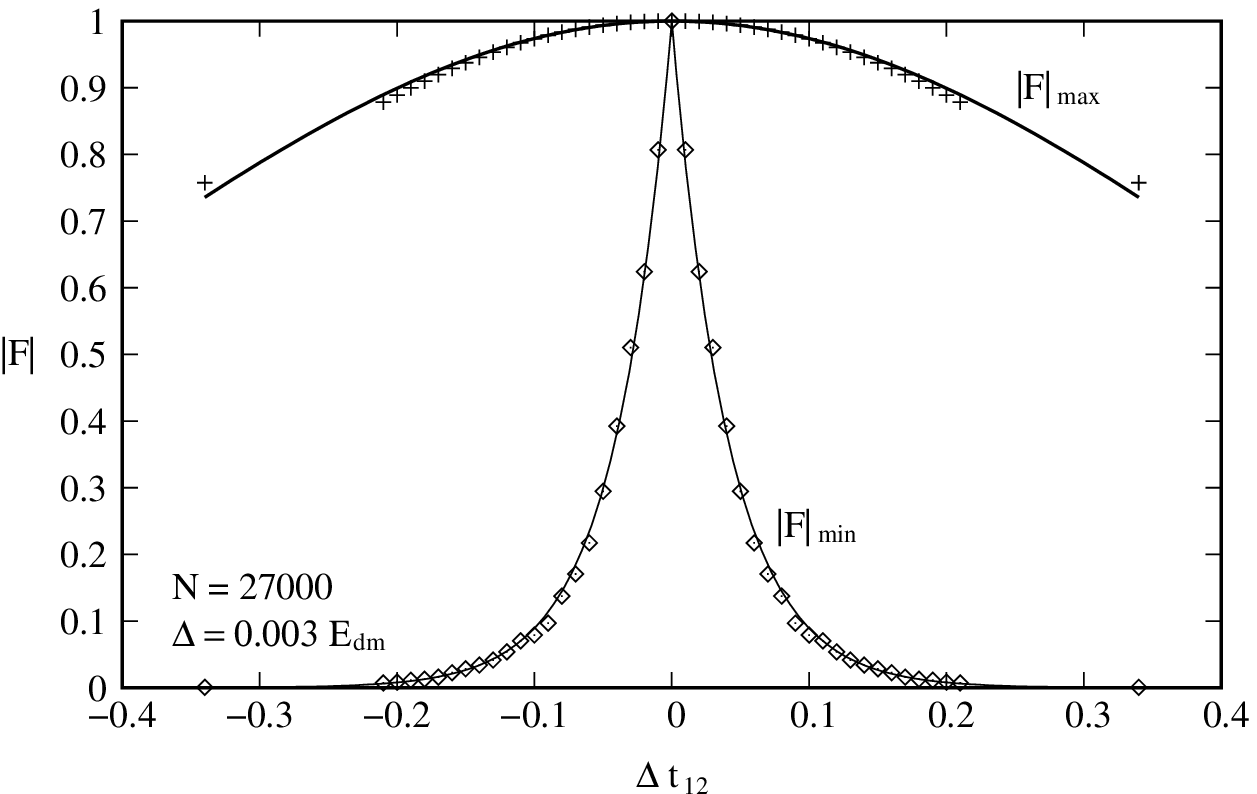}
\caption{\label{Dta0p3_27000.eps}
Same as \fno{Dta0p1_27000.eps} but with $\Dta = 0.003 E_{dm}$.}
\end{figure}
\begin{figure}
\includegraphics{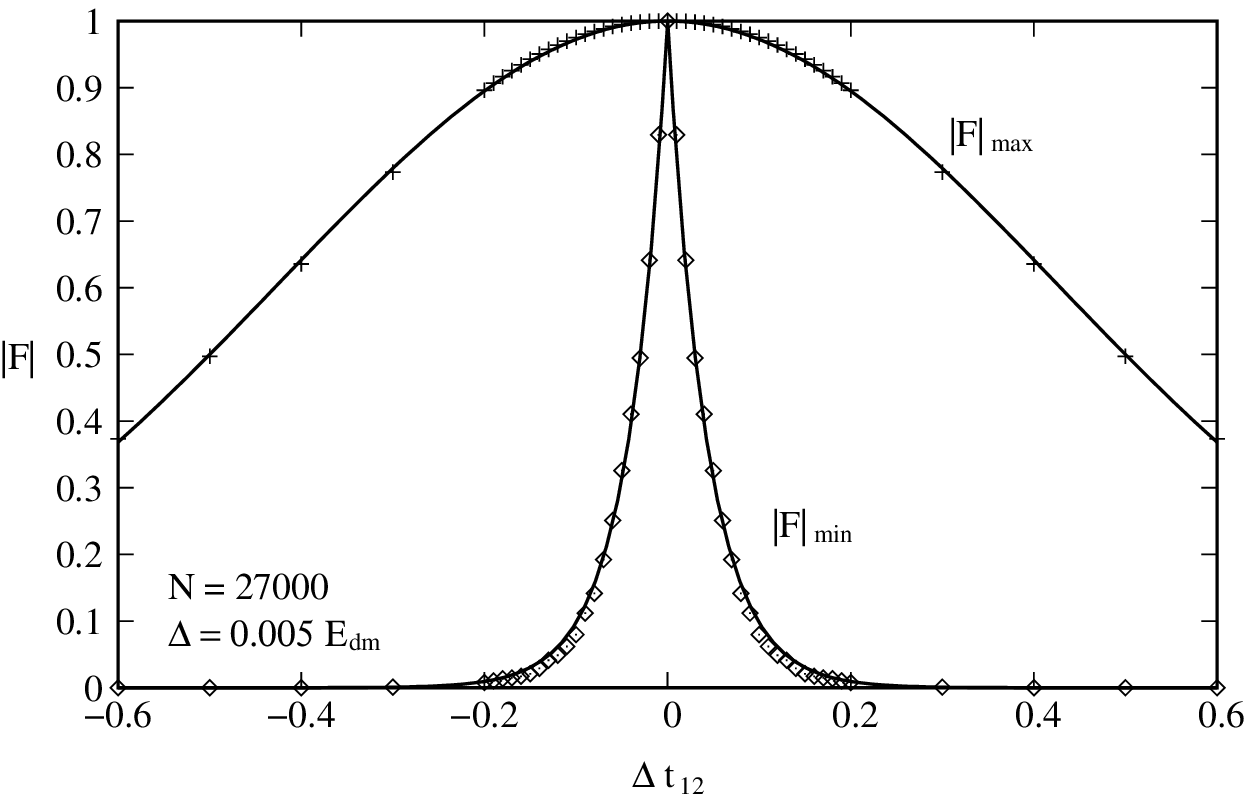}
\caption{\label{Dta0p5_27000.eps}
Same as \fno{Dta0p1_27000.eps} but with $\Dta = 0.005 E_{dm}$.}
\end{figure}


\begin{thebibliography}{99}

\bibitem{gsvbook} A comprehensive and lucid review of the entire field
of SMM's is given by D.~Gatteschi, R.~Sessoli, and J.~Villain,
{\it Molecular Nanomagnets\/} (Oxford University Press, Oxford, 2006).

\bibitem{hys9597}
C.~Paulsen and J.-G. Park, in {\it Quantum Tunneling of Magnetization%
---QTM '94\/}, edited by L.~Gunther and B.~Barbara (Kluwer, Dordecht,
1995);
%
J.~R. Friedman, M.~P. Sarachik, J.~Tejada, and R.~Ziolo,
Phys. Rev. Lett. {\bf 76}, 3830 (1996);
%
L.~Thomas et al.,
Nature (London) {\bf 383}, 145 (1996);
%
C.~Sangregorio et al.,
Phys. Rev. Lett. {\bf 78}, 4645 (1997).
%
Many more examples can be found throughout Ref.~\cite{gsvbook}.
%

\bibitem{phon} For excited states, such processes are important, and
have been studied extensively in \Mn12. For this, see
%
%
Chapter 10 of Ref.~\cite{gsvbook}, and references therein.

\bibitem{ww1}
W.~Wernsdorfer and R.~Sessoli, Science {\bf 284}, 133 (1999).

\bibitem{ww2} W.~Wernsdorfer, R.~Sessoli, A.~Caneschi, D.~Gatteschi,
A.~Cornia, and D.~Mailly, J. Appl. Phys. {\bf 87}, 5481 (2000).

\bibitem{ww3} W.~Wernsdorfer, R.~Sessoli, A.~Caneschi, D.~Gatteschi,
and A.~Cornia, Europhys. Lett. {\bf 50}, 552 (2000).

\bibitem{wer02}
W.~Wernsdorfer, N.~Aliaga-Alcalde, D.~Hendrickson, and G.~Christou,
Nature (London) {\bf 416}, 406 (2002).

\bibitem{edb02}
E.~del Barco et al.,
Europhys. Lett. {\bf 60}, 768--774 (2002).

\bibitem{lzs32}
L.~Landau, Phys. J. Sowjetunion {\bf 2}, 46 (1932);
C.~Zener, Proc. Roy. Soc. London A {\bf 137}, 696 (1932);
E.~C.~G. St\"uckelberg, Helv. Phys. Acta {\bf 5}, 369 (1932).

\bibitem{ag93}
A.~Garg, Europhys. Lett. {\bf 22}, 205 (1993).

\bibitem{ag9395}
A.~Garg, Phys. Rev. Lett. {\bf 70}, 1541 (1993);
{\it ibid\/} {\bf 74}, 1458 (1995).

\bibitem{pro9698}
N.~V. Prokofev and P.~Stamp,
J. Low. Temp. Phys. {\bf 104}, 143 (1996);
Phys. Rev. Lett. {\bf 80}, 5794 (1998).

\bibitem{sin03}
N.~A. Sinitsyn and N.~V. Prokofeev,
Phys. Rev. B {\bf 67}, 134403 (2003).

\bibitem{sin04}
N.~A. Sinitsyn and V.~V. Dobrovitski,
Phys. Rev. B {\bf 70}, 174449 (2004).

\bibitem{Cuc99} A.~Cuccoli, A.~Fort, A.~Rettori, E.~Adam, and
J.~Villain, Euro. Phys. J. B {\bf 12}, 39 (1999).

\bibitem{Fer03} J.~F. Fernandez and J.~J. Alonso,
Phys. Rev. Lett. {\bf 91}, 047202 (2003).

\bibitem{Kaya} Y.~Kayanuma, J. Phys. Soc. Jpn. {\bf 53},
108 (1984).

\bibitem{sixman}
A.~J. Leggett et al.,
Rev. Mod. Phys. {\bf 59}, 1 (1987).

\bibitem{am}
The sum of the inverse sixth power of the distance from a Bravais
lattice point to all other lattice points is tabulated for the
three cubic lattices in N.~W. Ashcroft and
N.~D. Mermin, {\it Solid State Physics\/} (Holt, Rinehart, and
Winston, New York, 1976), Table 20.2.

\bibitem{dob} W.~Zhang, N.~Konstantinidis, K.~A. Al-Hassanieh,
and V.~V. Dobrovitski, J. Phys. Cond. Matt. {\bf 19}, 083202
(2007). See Sec. 3.1.1. We are indebted to Dr. Dobrovitski for
telling us of this method.

\bibitem{B_clar} Note that $B_n$ entails the sum of the magnitudes
$h_i$, and not the magnitude of the vector sum $\sum_i \bh_i s_i$.

\bibitem{GR} I.~S. Gradshteyn and I.~M. Ryzhik, {\it Tables of
Integrals, Series, and Products\/}, corrected and enlarged edition
(Academic, New York, 1980). See formula 3.753.3.

\bibitem{ohm}
T.~Ohm, C.~Sangregorio, C.~Paulsen, Euro.~Phys.~J.~B~{\bf 6}, 195
(1998).

\bibitem{ww4} W.~Wernsdorfer, T.~Ohm, C.~Sangregorio, R.~Sessoli,
D.~Mailly, and C.~Paulsen, Phys. Rev. Lett. {\bf 82}, 3903 (1999).

\bibitem{Mukhin01} A.~Mukhin, B.~Gorshunov, M.~Dressel,
C.~Sangregorio, and D.~Gatteschi, Phys. Rev. B {\bf 63}, 214411
(2001).

\bibitem{Ber96} D.~V. Berkov, Phys. Rev. B {\bf 53}, 731 (1996).

\bibitem{somz} More precisely, it is conceivable that $\eta_i$
could vanish for isolated unequal values of $t_1$ and $t_2$,
but this would require a coincidence of the directions $\nhat_1$
and $\nhat_2$ defined in \eno{coh1}. Such a coincidence can
only happen for a measure-zero set of $K_i$, $\eps_i$, $t_1$,
and $t_2$. It can therefore be regarded as accidental, and
for all practical purposes we can say that $\eta_i > 0$
if $t_{12} \ne 0$.

\bibitem{kg07}
E. Ke\c{c}ecio\u{g}lu and A.~Garg, Phys. Rev. B {\bf 76},
134405 (2007).

\bibitem{merci} We were led to think about the points in this
paragraph in a correspondence with Jacques Villain and Wolfgang
Wernsdorfer.

\bibitem{werns05}
W.~Wernsdorfer, S. Bhaduri, A. Vinslava, and G. Christou,
Phys. Rev. B {\bf 72}, 214429 (2005).

\bibitem{werns08} W.~Wernsdorfer, http://arxiv.org/abs/0804.1246.v1.

\bibitem{quad_Fmax} Even though we do not need it to find
$\sum_i \eta_i$, the argument that follows provides still more
support for our approach. The $\exp(-b t_{12}^2)$ behaviour of
$|F|_{\rm max}$ can be understood if we note
that the weight of the delta functions in the replacements
(\ref{S1_dta}) and (\ref{S3_dta}) should be multiplied by another factor
$(1 - \Dta |t_{12}|/\pi)$, because the integrals over
$\Oimi$ begin from $\Dta$, not $0$. (This is readily seen by
performing the integrals from $0$ to $\infty$, and from 0 to
$\Dta$, and subtracting.) When we make these changes, and fix $c$
according to \eno{fix_c}, we find
$S_1 + S_2 - S_3 = b |\Dta t_{12}|^2$, with $ b = 2.02 E_{dm}/E_b$,
or just 2.02 if $E_{dm} = E_b$. Our numerical fits yield
$b \sim 2.1$--$2.8$, which is close enough given the crudeness of
our estimates.

\end{thebibliography}
\end{document}